\documentclass[psfig]{article}
\usepackage{supertabular}
\usepackage{subfigure}
\usepackage{lscape}
\usepackage{longtable}
\usepackage{psfig}
\usepackage{rotating}
\usepackage{graphicx}
\usepackage{rotating}
\begin{document}
\textheight=690pt
\textwidth=480pt
\oddsidemargin=-20pt
\topmargin=-20pt
\footskip=0pt
\marginparwidth=0pt

\title{The ATCA Intraday Variability Survey of Extragalactic Radio Sources}
\author{L. L. Kedziora-Chudczer,$^{1,2}$ \footnote{Present address: Anglo-Australian Observatory, PO Box 296, Epping 1710} D. L. Jauncey,$^{2}$ M. H. Wieringa,$^{2}$ \\ A. K. Tzioumis$^{2}$ and J. E. Reynolds$^{2}$\\
 $^{1}$SRCfTA, The University of Sydney, Australia\\
$^{2}$ ATNF, CSIRO, Australia}
\maketitle
\begin{abstract}
We present the results of an Australia Telescope Compact Array (ATCA) 
survey for intra-day variability (IDV) of the total and polarized flux 
densities of 118 compact, flat spectrum, extragalactic radio sources 
from the Parkes 2.7 GHz Survey. 
A total of 22 total flux density IDV sources were discovered and 15 
sources were found to show IDV of their polarized flux density.
We discuss the statistical properties of the IDV sources, including 
the distribution of source modulation indices, and the dependence of the 
variability amplitude on source spectral index and on Galactic position.
We suggest interstellar scintillation (ISS) in the Galactic 
interstellar medium as the most likely mechanism for IDV. Even so, the 
inferred high brightness temperatures cannot be easily explained.  \end{abstract}
Keywords: \
active galactic nuclei: blazars and intraday variability --
interstellar medium: interstellar scintillation

\section{Introduction}
The first systematic investigation of short time scale variability in a 
large sample of extragalactic radio sources was carried out by Heeschen (1984). 
He observed a sample of 226 strong sources ($S_{6cm}\ge 1.0$Jy) at 9 cm.
 Most of the compact, flat spectrum sources in his sample showed significant 
flux density fluctuations, with an average amplitude of 1.5\% on time scales 
of 2-20 days. The steep spectrum radio sources did not appear to vary. This, 
so called {\em flickering} was interpreted in terms of refractive scintillations 
caused by the inhomogeneous structure of the interstellar medium in our Galaxy.

A search for more
rapid and possibly stronger flickering through better sampled observations of 
compact, flat spectrum radio sources was undertaken by Witzel et al. (1986) and Heeschen et al. (1987)  
at the 100m radio telescope of the Max Plank Institut f\"ur Radioastronomie (MPIfR) . 
They reported 10\% flux density variations at 11 cm in the BL Lac object 
S5~0716+71 on a time scale of a day, and  they also confirmed the 
statistical difference in the variability of steep and flat-spectrum radio sources. 

Multi-frequency observations of intraday variable sources at MPIfR and the VLA  
suggested that the variability tends to be strongest at 2.7 GHz and well 
correlated between 8 and 2.7 GHz (Wagner \& Witzel 1995).
Monitoring of the polarized flux density for the two strongest IDV sources, S5~0716+714 and
S5~0917+624, revealed polarization variability correlated with the total
flux density fluctuations for S5~0716+714 and anti-correlated for S5~0917+624 
(Wegner 1994). Rapid variability in the position angle of the polarized flux 
density was also reported in these sources (Quirrenbach et al. 1989).

The interest in IDV at cm wavelengths is motivated by the high brightness 
temperatures inferred from the variability timescales, well above the 
$T_{C}\sim 10^{12}$ K inverse Compton limit (Kellermann \& Pauliny-Toth 1969). 
Possible mechanisms which can lead to IDV fall into two categories, those 
intrinsic to the source (e.g. shocks in jets, coherent emission) and 
propagation effects, such as microlensing and ISS (see e.g. the review by 
Wagner \& Witzel 1995). None of the models proposed has been successful in 
providing a complete description of the observed properties of this phenomenon. 
It is likely that both extrinsic and intrinsic mechanisms are 
responsible for the observed rapid flux density fluctuations at cm wavelengths. 
The present ATCA IDV Survey and monitoring program provides a large, uniform, high precision, multi-wavelength set of observations, which aims to describe the common properties of IDV sources and shed the light at the origin of short term variations.

\section{The sample selection}
\label{sect1}

The sample selected for the ATCA survey is drawn from the Parkes-Tidbinbilla 
Interferometer (PTI) flat spectrum ($\alpha^{2.7GHz}_{5.0GHz} > -0.5$)\footnote{Hereafter, the spectral index is 
defined as $S\propto \nu ^{\alpha }$ where
$\nu $ is the frequency of radiation and $S$ is the flux density.} survey  
(Duncan et al. 1993). Sources were chosen which had a ratio of the PTI flux 
density to total flux density larger than 0.9.
Flat-spectrum, compact
components of radio sources are usually self-absorbed and coincide with the 
optical quasar or AGN (e.g. Jauncey et al. 1982). We chose such sources since 
(a) monitoring of flux density with an interferometer such as the ATCA can be done faster and more accurately for unresolved sources, and (b) it is known from previous studies (e.g. Heeschen et al. 1987) that flat-spectrum 
sources are more likely to show IDV than steep-spectrum sources.  

The final sample contains 118 sources, randomly distributed over the southern sky for $b<10^{\circ} $ (Figure~\ref{fig29}). They are listed in
Table~\ref{tab:1} along with identifications, galactic coordinates, redshift and measured source structure information. The sample 
contains 90 (76\%) quasars, 13 (11\%) BL Lacs, 7 (6\%) galaxies and 8 (7\%) 
unidentified sources.

\section{Observations}
\label{sect2}
The main objective of the Survey was to obtain well sampled, multi-frequency total and polarized 
flux density measurements for all sources in the sample within the 96 hours 
scheduled for each observing run. Our observations were carried out at the ATCA, 
an east-west earth-rotation synthesis interferometer, with six 22 metre antennas 
on a 6 km baseline (Sinclair \& Gough 1991). The ATCA has a number of desirable features 
for a study of short timescale variability:
\begin{itemize}
\item[(i)] short integration time which enables frequent sampling,
\item[(ii)]  broadband instantaneous frequency coverage,
\item[(iii)]  the high accuracy and sensitivity essential to monitor 
polarization and to explore the low amplitude, `flicker' of extragalactic sources.   
\end{itemize}
During the two observing sessions in May and August 1994, the flux 
densities of sources in the sample were measured every 2 hours at the four ATCA frequencies, 
1.4, 2.4, 4.8 and 8.6 GHz. We used PKS~1934-638 as the primary flux density calibrator. Additionally the known steep spectrum sources PKS~0518+165, PKS~0134+329, PKS~1328+307 and PKS~0823$-$500 were observed throughout each session to check the overall gain stability and calibration.

The observations were performed in continuum mode with 128 MHz bandwidth 
divided into 32 channels. This configuration enables measurements of four 
polarization products for each frequency. The sampling time was chosen to be 
1 minute, which is equivalent to six 10 s integrations on source. To avoid long slew times, 
the sources were observed in two main groups, either north or south of zenith. 
Each group was observed over 24 hours with each source being observed every 2 hours 
while it was above 20 degrees elevation. The flux density of each source, 
including calibrators, was measured typically 6 times over 12 hours, and the 
measurements repeated in the same way 36 hours later.

The gain-stability was monitored by regular observations of PKS~1934-638 and the other non-variable sources, and so the scatter in their measured flux densities is an estimate of the uncertainties in the individual flux density measurements. 
This includes the effects of antenna pointing errors and gain changes with 
elevation and time. 

The telescope pointing, typically 7 arcseconds rms, was determined before each 
observing session. The positions of all sample sources are known to an accuracy 
of 0.5 arcseconds or better, so that flux density errors due to miss-pointing do not exceed 0.1\% at the highest frequencies. 
The observed rms errors for the primary flux density calibrator, PKS 1934-638 
for an individual 1 minute observing scan are listed in Table~\ref{tab:9} for 
each frequency. Thus, at the flux density limit, $S\geq0.5$ Jy, the gain 
stability and pointing errors dominate, while the thermal noise is typically less than a mJy.  

The source structure information presented in columns 6-9 of Table~\ref{tab:1} is derived from the closure phase estimates (Jennison 1958) of the possible extended component in the measured flux density given by:
\begin{equation}  
S_{ext}\, = C\, \overline{S}\, (\sigma ^{2}_{obs}-\sigma^{2}_{theo})^{1/2},
\label{eq1}
\end{equation}
where $\overline{S}$ is the flux density averaged over 12 hours, 
$\sigma_{obs}$ is the observed rms of closure phases measured for all independent
baselines and $\sigma_{theo}$ is a theoretically predicted rms of closure phases for a point source calculated by use of calibration from antenna noise diodes (Sault 1994). $C$ is a constant of the order of unity.  

We find that most of the target sources of the ATCA IDV Survey are unresolved at 8.6 and 4.8 GHz. In particular, 72\% of the sample sources show less than 1\% extended flux density at 4.8 GHz. However only 22\% and 5\% of the sources at 2.4 and 1.4 GHz respectively satisfy this condition (Table~\ref{tab:1}). 

Almost half of the sample (62 sources) shows more than 5\% flux density in an extended component at 1.4 GHz. The effects of weak extended structure and confusing sources in the primary beam result in much larger object-dependent flux density errors at this frequency. Therefore the results of the 1.4 GHz measurements are excluded from the statistical discussion of the survey. 
\section{Results}
\label{sect3}
\subsection{Flux densities}
The results of the Survey are presented in 
Table~\ref{tab:4} and Table~\ref{tab:5}, which show the averaged total ($\overline{S}$) 
and polarized ($\overline{P}$) flux densities, averaged polarization position angles ($\psi$) and the corresponding modulation indices ($\mu_{\overline{S}}$, $\mu_{\overline{P}}$).  

The total flux densities in Table~\ref{tab:4} are the averages of all 1-minute 
measurements within a given observing session over typically 96 hours.  
In each session the number of data points, $N$ used in the averages, ranges 
between 5 and 36 with a median of 11. Each data-point used to obtain the flux density in Table~\ref{tab:4} is an average over all ATCA baselines of typically six 10-sec integrations.

Seven sources (PKS~0023$-$263, PKS~0122$-$003, PKS~0138$-$097, PKS~208$-$512, 
PKS~0454$-$463, PKS~1253$-$055 and PKS~2052$-$474) were excluded from the 
sample because of the presence of either extended structure at all frequencies 
or strong confusing sources in the primary beam. 
A further 36 sources were found to be confused, or in a few cases, extended 
at 2.4 and 1.4 GHz. These have blank entries  in Tables~\ref{tab:4} and~\ref{tab:5} at the corresponding frequencies. In this 
way the consistently high-quality data at high frequencies was maintained by 
reducing the size of the sample slightly. The final numbers of sources 
at each frequency in the Table~\ref{tab:4} are 111 for 8.6 GHz, 105 for 
4.8 GHz, 74 for 2.4 GHz and 55 for 1.4 GHz. 

The 1.4 GHz data was not used in the IDV investigation, although the 
average flux densities have been retained in both Tables for the sources with $S_{ext}\geq 3.5\%$.

The errors in the observed flux densities in Tables~\ref{tab:4} and ~\ref{tab:5} have 
been determined from the observed scatter in the strong sources. Table~\ref {tab:9} shows the observed scatter for the primary flux density calibrator PKS~1934$-$638, the non-variable point source most frequently observed throughout each Survey session. The measured rms noise integrated over 1 minute never exceeds 4 mJy. With normalized flux density errors of no more than $0.5\% $ at each of the four frequencies, the present accuracy compares favourably with the precision of single dish measurements.

\begin{figure*}
\begin{center}
\psfig{bbllx=0pt,bblly=0pt,bburx=695pt,bbury=475pt,file=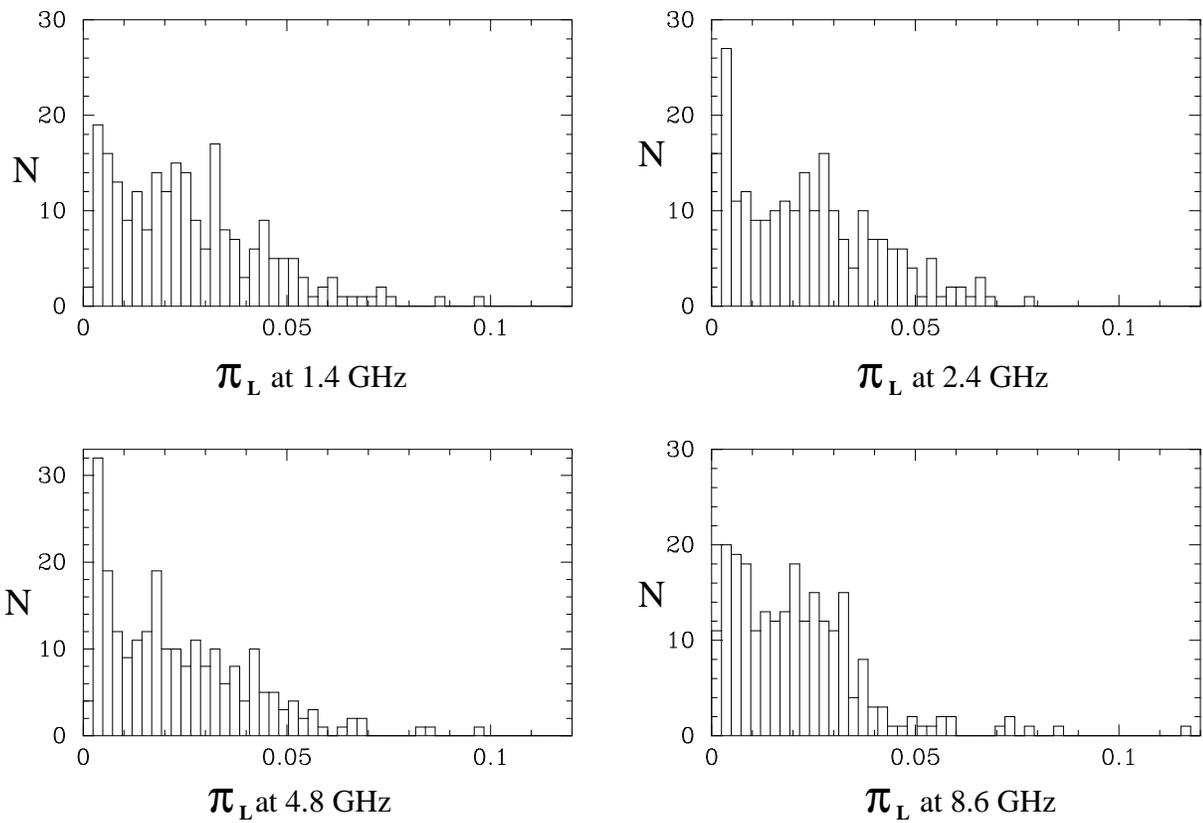,clip=,angle=0,height=110mm,width=160mm} 
\caption{Distribution of the fractional linear polarization averaged over a duration of each observing session (May and August 1994 combined) for all sources in the sample.}
\label{fig17}
\end{center}
\end{figure*}

The ATCA antennas each have two, on-axis, orthogonal, linearly polarized feeds at all frequencies, which enable easy acquisition of all Stokes parameters. The polarized flux densities $P$, and the position angles, $\psi $, along with the modulation indices, $\mu_{P}$, are presented for the 74 sources with $P > 15$ mJy in Table~\ref{tab:5}. 
The degree of linear polarization observed for compact, flat-spectrum, synchrotron sources is typically a small fraction of the total flux density ($\leq 5\% $) (Aller at al. 1985), and this is the case for the majority of sources observed in the present Survey (Figure~\ref{fig17}). 

Monitoring sessions in May and August 1994 provided two datasets of the same sample. To determine if the data obtained in May and August 1994 is comparable in the statistical sense, we applied the Kolmogorov-Smirnov test for the source number distributions of normalized standard deviations or modulation indices, $\mu $ ($=\sigma_{S}/\overline{S}$) of the flux density measurements. This found no statistically significant difference between the May and August 1994 results at the confidence level 0.05\%.

\subsection{Spectral index properties}
\label{spec}
The average spectral indices: $\alpha^{8.6}_{4.8}$, $\alpha^{4.8}_{2.4}$, $\alpha^{2.4}_{1.4}$ and $\alpha^{8.6}_{1.4}$ are listed in Table~\ref{tab:6}. The uncertainty of the spectral index measurement depends on the modulation indices at the relevant frequencies and on the ratio of these frequencies. The estimated $\sigma_{\alpha}\sim2\%$ for non-variable sources is slightly dependent on the frequency pair used.

\begin{figure}
\centering{
\psfig{bbllx=0pt,bblly=0pt,bburx=705pt,bbury=490pt,file=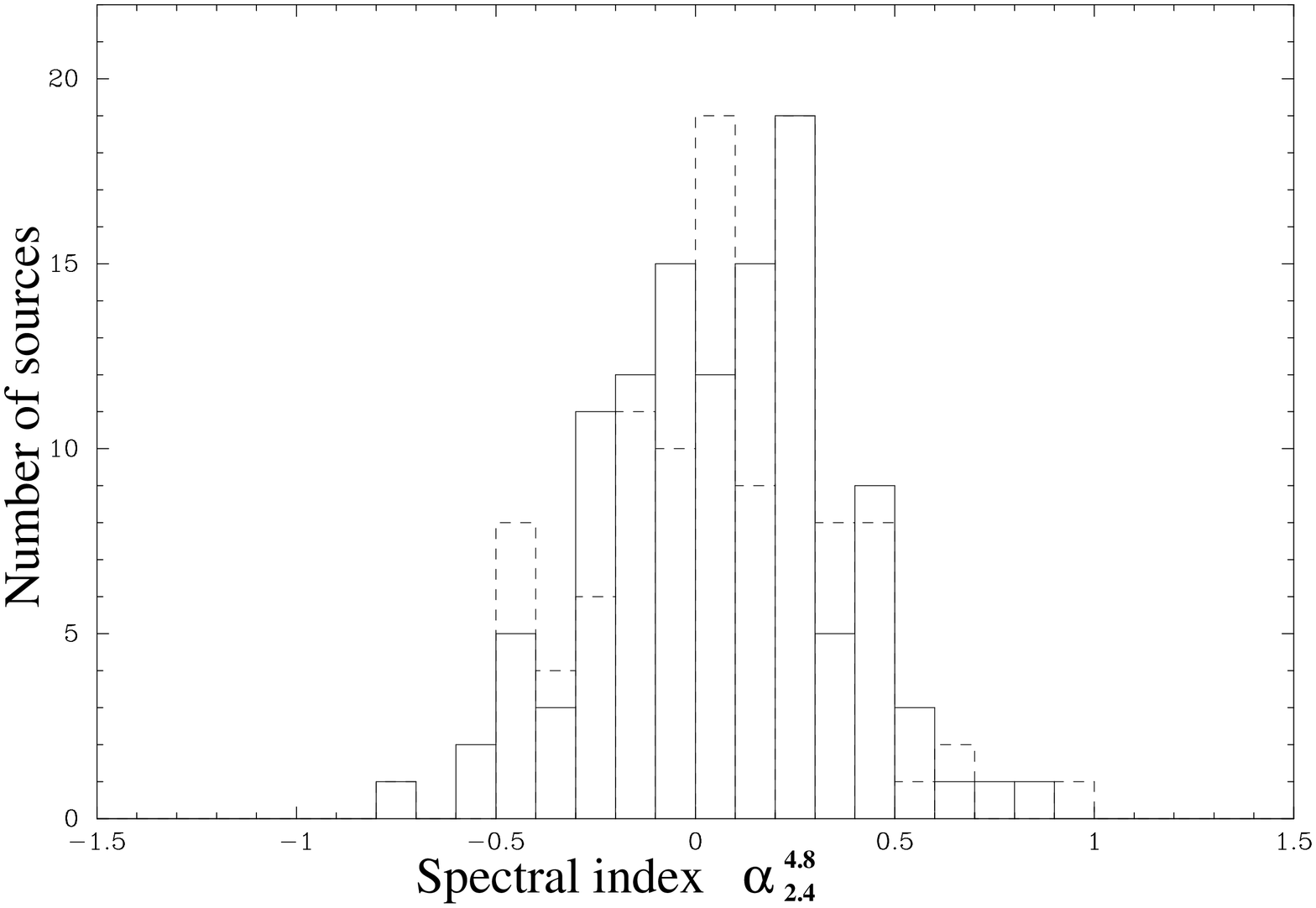,angle=0,height=66mm,width=90mm}
\vspace{5mm}
\psfig{bbllx=0pt,bblly=0pt,bburx=700pt,bbury=530pt,file=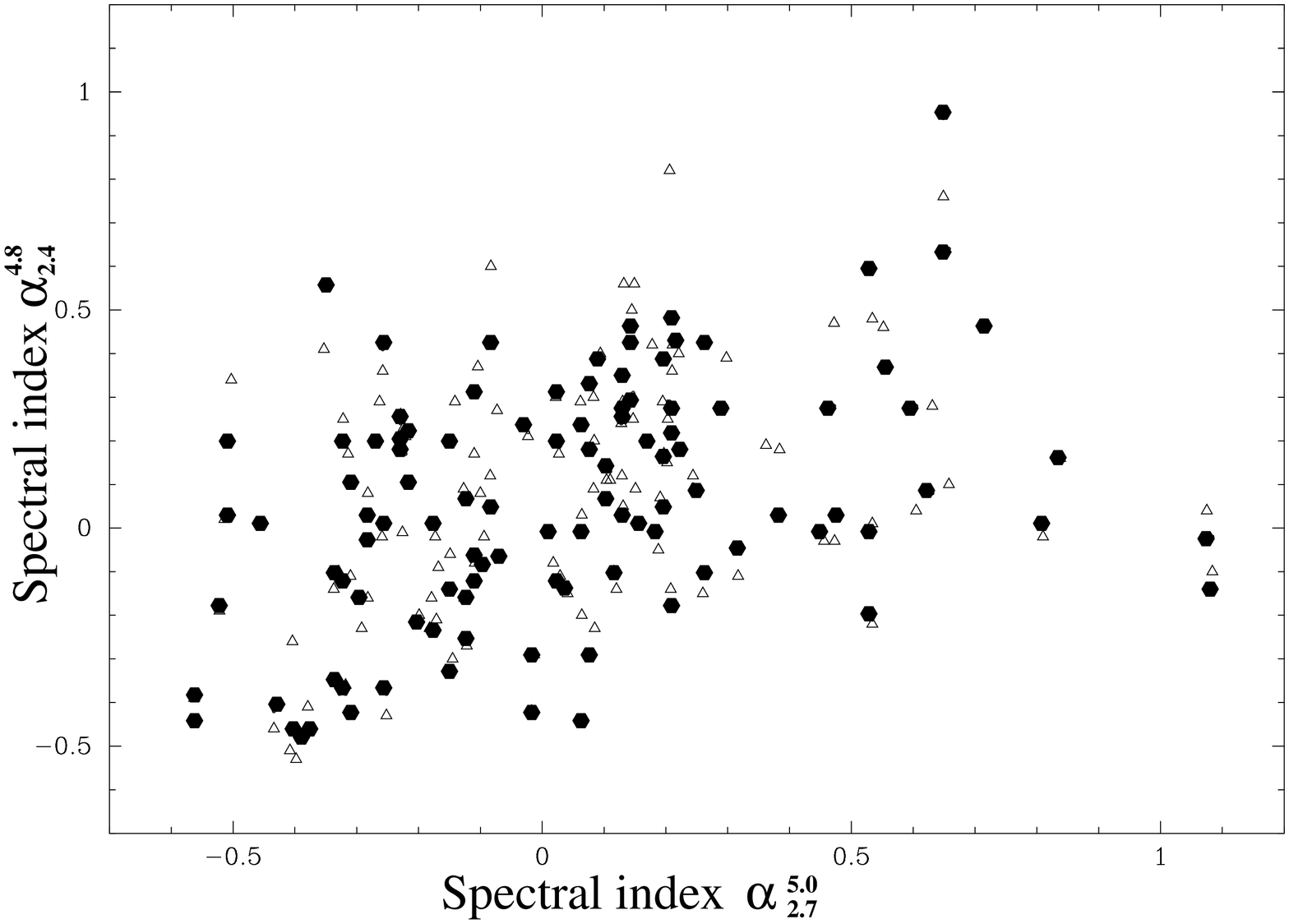,angle=0,height=70mm,width=89mm}}
\caption{Upper panel: the number distribution of the Survey spectral indices $\alpha
_{2.4}^{4.8}$ of all compact sources for May (full line) and August 1994 (dashed line). Lower panel: The scatter plot of Parkes spectral indices $\alpha _{2.7}^{5.0}$ versus spectral indices $\alpha
_{2.4}^{4.8}$ of all sources for May (hexagons) August 1994 (triangles).}
\label{fig15}
\end{figure}

The distribution of the spectral indices $\alpha_{2.4}^{4.8}$ is shown in Figure~\ref{fig15}. There are only three sources with $\alpha_{2.4}^{4.8}\leq -0.5$. If spectral indices $\alpha_{2.4}^{4.8}$ are compared with indices $\alpha_{2.7}^{5.0}$ measured with the Parkes telescope, it is evident
that, while the sources may vary considerably, very few, if any of the original flat-spectrum sources change to steep-spectrum sources (Figure~\ref{fig15}(b)).
The differences in the distribution of spectral indices between May and August 1994 (Figure~\ref{fig15}(a)) reflect the level of monthly variability of the flat-spectrum sources. 

\begin{figure*}
\centering
\psfig{bbllx=0pt,bblly=0pt,bburx=710pt,bbury=1545pt,file=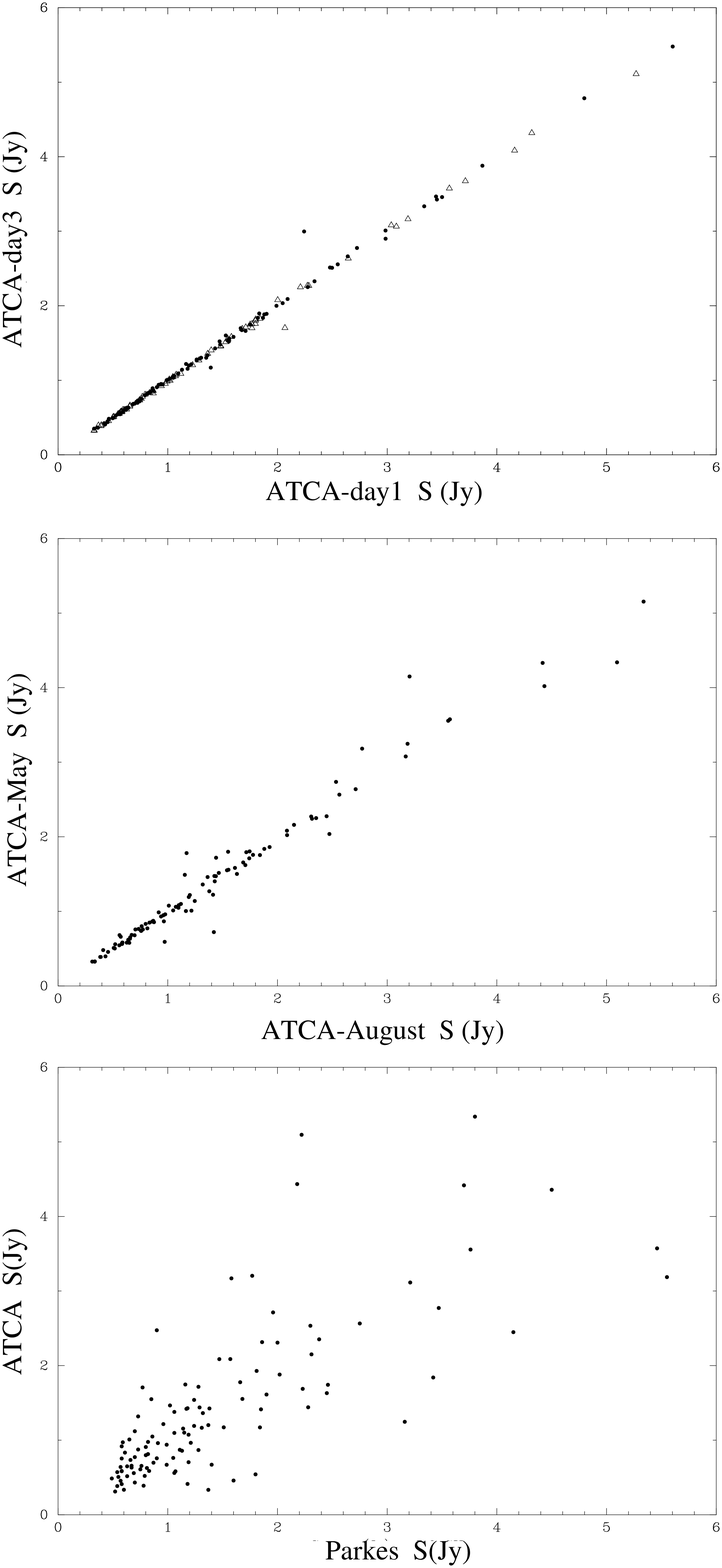,clip=,angle=0,height=200mm,width=0.50\textwidth}
\caption{Comparison of flux densities at 4.8 GHz for progressively longer time separations. Top: the measurements taken one day apart with the ATCA, dots - May data, triangles - August 1994 data. Middle: the average flux densities measured 3 months apart with the ATCA. Bottom: the measurements with ATCA in May 1994 plotted against flux densities listed in the Parkes catalogue. }
\label{fig155}
\end{figure*}

The flat-spectrum sources ($\alpha_{2.7}^{5.0}\geq-0.5$) remain flat-spectrum over much longer timescales, which is illustrated in the scatter plots of average flux densities shown for the increasing time difference between measurements (Figure~\ref{fig155}).   

The shapes of our four-frequency spectra can be described as: decreasing with frequency, straight, peaked, concave, inverted and complex. The largest population represented in the Survey sample (42 sources corresponding to 36\%) are peaked spectrum sources. The second largest groups are 22 (19\%) decreasing and 21 (18\%) inverted spectrum sources. There are 3 (2.5\%) sources with a straight spectrum (with $< 50$ mJy deviations from an $\alpha =0$ line), 11 (9.3\%) concave spectrum and 1 (0.8\%) complex spectrum source.

The shape of the spectrum changed for the 15\% of sources between two observing sessions. Interestingly, no transition from peaked to concave spectrum was observed. This may suggest that, at the typical timescale of months,  the flux density changes are due to the slower `van der Laan - type' evolution corresponding to the physical expansion of the region of the original outburst (van der Laan 1966). 
\subsection{Short-term variability}
\label{sect4}
The normalized amplitude of variability is expressed in terms of a modulation index, $\mu $,  calculated over 96 hours, the duration of each observing session.   
Figure~\ref{fig155} illustrates the different timescales of flux density variations in our sample by comparing measurements obtained one day apart, 3 months apart and a few decades apart. While large flux density fluctuations over the short (day-to-day) timescales are rare in an ensemble of flat-spectrum, compact sources it is clear that all these sources vary significantly on a timescale of decades.

\begin{figure*}
\centering
\psfig{bbllx=563pt,bblly=100pt,bburx=93pt,bbury=799pt,file=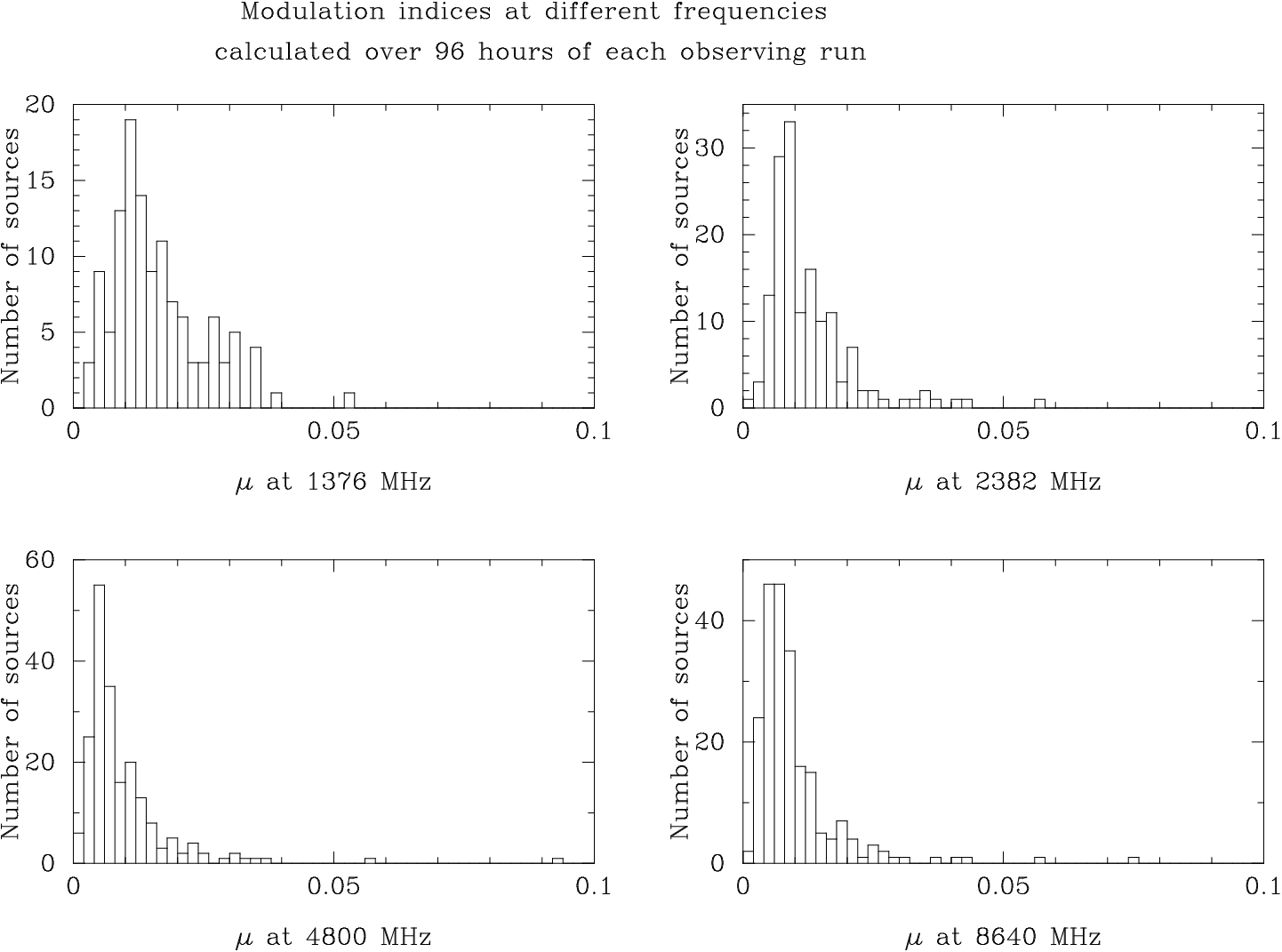,clip=,angle=270,height=120mm,width=180mm}
\caption{The combined distribution of modulation indices over two observing sessions, May and August 1994 for all frequencies.}
\label{fig21a}
\end{figure*}

The number distribution of modulation indices is shown in Figure~\ref{fig21a}.
The peaks in these distributions give an independent estimate of the normalized rms error in our measurements of flux densities. If one approximates each distribution by a Gaussian plus a tail of high modulation indices, the typical measurement rms values for the frequencies 8.6, 4.8, 2.4 and 1.4 GHz are 0.65, 0.6, 0.8 and 1.2\% respectively.
The increased rms distribution of modulation indices at 1.4 GHz is due to confusion and weak extended components present in the sources (as noted in  section~\ref {sect2}).   

Those sources found in the tails of modulation indices distributions at different frequencies (Figure~\ref{fig21a}) are the sources, which exhibit significant IDV.
There are 22 sources (19\% of the total sample) which show modulation indices larger than $3\sigma $ at any frequency (8.6, 4.8 and 2.4 GHz) during either the  May or August observing sessions. They are listed in Table~\ref{tab:11}.

In Table~\ref{tab:11} there are 43 entries with $\mu > 3\sigma $ typed in bold. If these were purely random events, rather than any real effect, one would expect to see 4 sources showing two entries with $\mu > 3\sigma $, and no sources with three or more entries. Yet there are 6 sources with two entries, and 6 sources with three entries, and one source, PKS~1519-273, with four entries. Such a high incidence of multiple entries suggests that the results presented in Table~\ref{tab:11} represent a real effect.

Out of these 22 IDVs, 10 (45\%) have a peaked spectrum, and 8 (36\%) have an inverted spectrum. The highest incidence of IDV occurs for inverted spectrum sources (39\% of all inverted spectrum sources in the sample), followed by peaked spectrum sources (24\%) and decreasing spectrum sources (9\%).
\subsection{Long-term variability ($\leq 3$ months)}
\label{sect4_2}
\begin{figure*}
\renewcommand{\subfigcapskip}{0pt}
\renewcommand{\subfigtopskip}{5pt}
\centering
\mbox{\subfigure[]{\psfig{bbllx=372pt,bblly=0pt,bburx=732pt,bbury=228pt,file=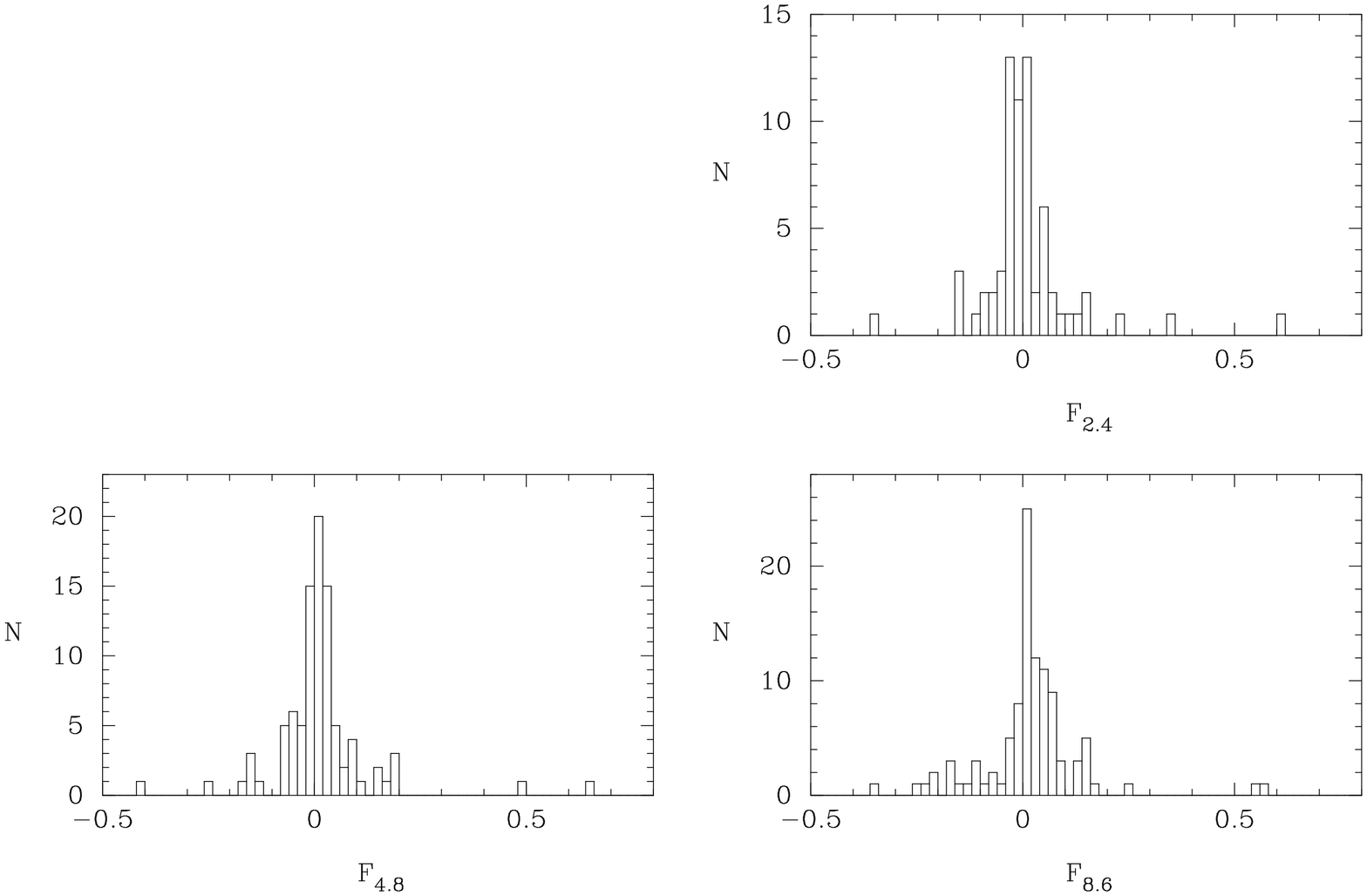,clip=,angle=0,height=50mm,width=0.48\textwidth}}\quad
\subfigure[]{\psfig{bbllx=372pt,bblly=0pt,bburx=732pt,bbury=228pt,file=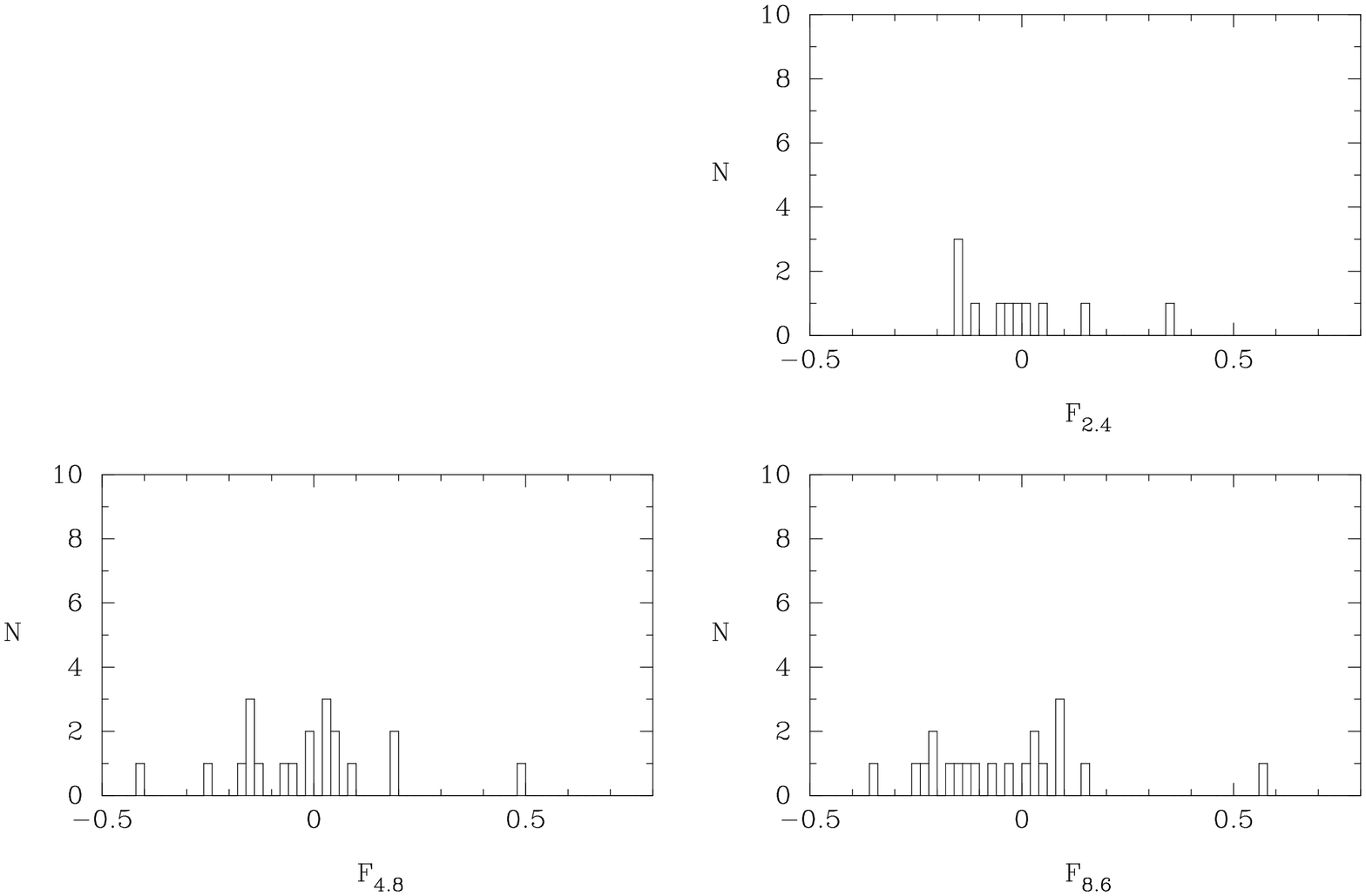,clip=,angle=0,height=50mm,width=0.48\textwidth}}}
\mbox{\subfigure[]{\psfig{bbllx=0pt,bblly=0pt,bburx=360pt,bbury=228pt,file=all_ave.ps,clip=,angle=0,height=50mm,width=0.48\textwidth}}\quad
\subfigure[]{\psfig{bbllx=0pt,bblly=0pt,bburx=360pt,bbury=228pt,file=idv_ave.ps,clip=,angle=0,height=50mm,width=0.48\textwidth}}}
\mbox{\subfigure[]{\psfig{bbllx=372pt,bblly=234pt,bburx=732pt,bbury=464pt,file=all_ave.ps,clip=,angle=0,height=50mm,width=0.48\textwidth}}\quad
\subfigure[]{\psfig{bbllx=372pt,bblly=234pt,bburx=732pt,bbury=464pt,file=idv_ave.ps,clip=,angle=0,height=50mm,width=0.48\textwidth}}}
\caption[Distribution of the fractional difference in average flux density between two observing sessions in May and August 1994.] {Distribution of the fractional difference in average flux density between two observing sessions: May and August 1994. Left panels: the whole sample. Right panels: only IDV sources from Table 6 are included.}
\label{fig26}
\end{figure*}

The distribution of the fractional difference, $F_{\nu }$, in the average flux density between May and August 1994 observations:
\begin{equation}
F_{\nu }=2\,\frac{\overline{S}^{May}_{\nu }-\overline{S}^{Aug}_{\nu }}{\overline{S}^{May}_{\nu }+\overline{S}^{Aug}_{\nu }}
\label{eq:f}
\end{equation} 
is shown in Figure~\ref{fig26}.
The measurement errors of the fractional difference are estimated from the rms of the flux density measurement distributions. The $3\sigma $ level is less than 4\% for each frequency.  We find $40\%$ sources which vary by more than 4\% at 4.8 and 2.4 GHz. The number of sources variable on monthly timescales becomes larger with increasing frequency.   

A substantial fraction of IDV sources are also long term variable (80\% at 8.6 GHz, 75\% at 4.8 GHz and 73\% at 2.4 GHz). This characteristic may prove useful in future searches for IDV sources. 

Out of the 22 IDV sources only six were found to show significant IDV in both observing sessions. The reminder of the sources (two thirds) has either a shorter duty cycle than $\sim 3$ months, or their variability decreases below our $3\sigma$ inclusion limit. PKS~0405$-$385 is the most spectacular example of the short lived rapid variability (Kedziora-Chudczer et al. 1997). Yet the other strong IDV sources (PKS~1034$-$293, PKS~1144$-$379 and PKS~1519$-$273) vary each time they were observed during the follow-ups of the Survey (Kedziora-Chudczer et al. 1998, 2000).

We also found that sources which show a difference in the shape of their total flux density spectrum between May and August sessions are twice as likely to show IDV as the sources without a significant spectral change.  As seen from Figure~\ref{fig26} these monthly spectral changes are not necessarily associated with the overall strong flux density change at all frequencies.
\subsection{Polarization variability}
\label{sect4_3}
\begin{figure}
\renewcommand{\subfigcapskip}{1pt}
\renewcommand{\subfigtopskip}{3pt}
\centering
\psfig{bbllx=372pt,bblly=0pt,bburx=732pt,bbury=228pt,file=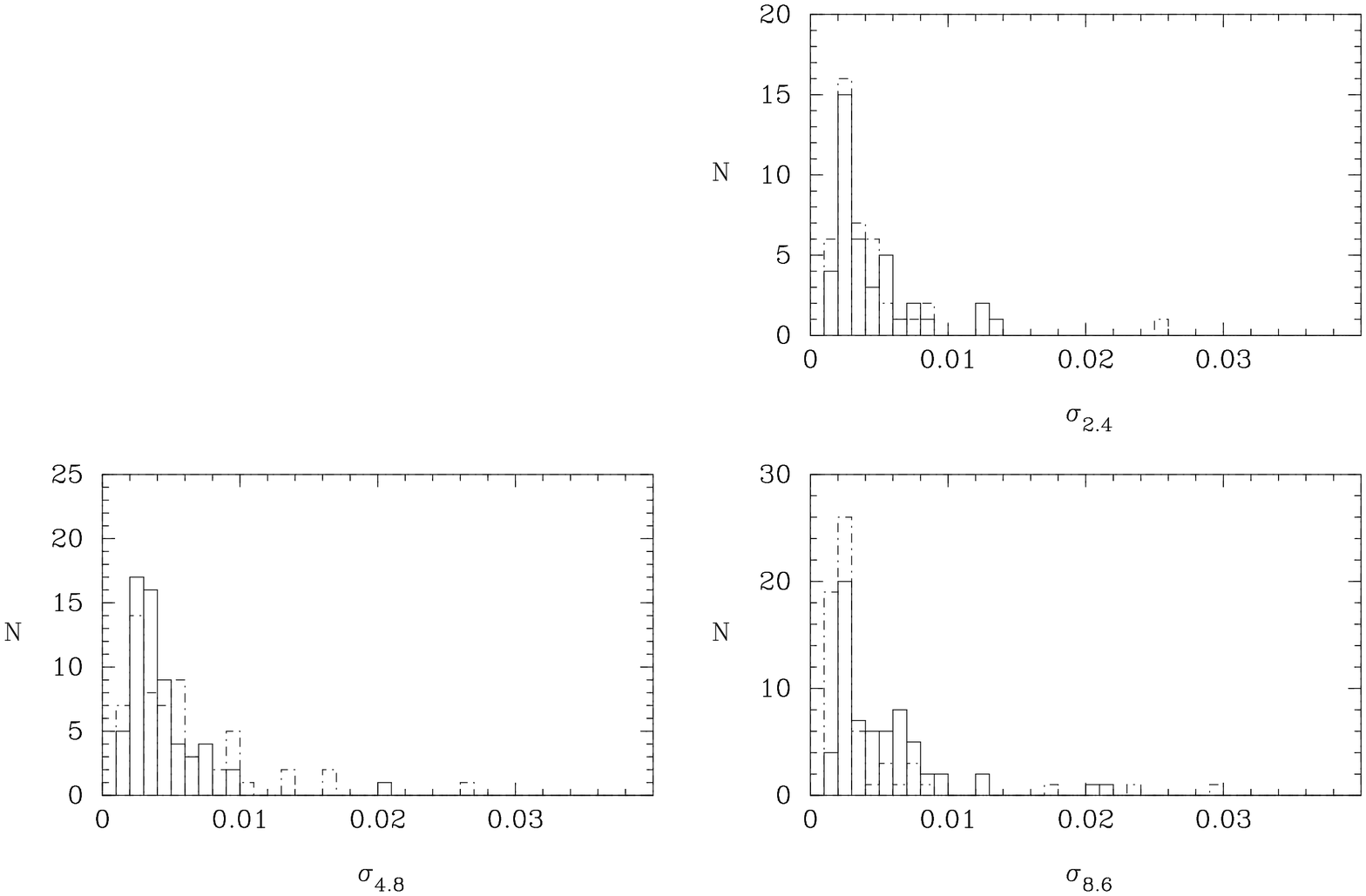,clip=,angle=0,height=50mm,width=0.48\textwidth}
\psfig{bbllx=0pt,bblly=0pt,bburx=360pt,bbury=228pt,file=pol2.ps,clip=,angle=0,height=50mm,width=0.48\textwidth}
\psfig{bbllx=372pt,bblly=234pt,bburx=732pt,bbury=464pt,file=pol2.ps,clip=,angle=0,height=50mm,width=0.48\textwidth}
\caption{The number distribution of the rms in polarized flux densities at 8.6, 4.8 and 2.3 GHz from top to bottom.}
\label{fig25}
\end{figure}

The polarization modulation index, $\mu _{P}$, is defined as the rms of the polarized flux density normalized by the average polarized flux density. 
For the ATCA the instrumental polarization can be easily calibrated, therefore the main contribution to observed measurement errors comes from the antenna gain stability. The distribution of measured rms for three frequencies 8.6, 4.8 and 2.4 GHz is presented in Figure~\ref{fig25}. An estimate of the measurement error, $\sigma=4$ mJy, was derived by finding the range of rms within which two thirds of the sources are distributed around the peak of the distribution. 
Only 6 (8\%) sources vary above $4\sigma$ at any frequency. These are PKS~0215+015, PKS~0537$-$441, PKS~0607$-$157, PKS~1144$-$379, PKS~1334$-$127 and PKS~1519$-$273. The most significant variability of polarized flux is seen for sources which already were identified as IDVs in the total flux density. The polarized flux density IDV of PKS~1334$-$127 is also independently found in VLBI data (Kochanev \& Gabuzda 1998).

We also analysed the day-to-day polarization variability by self-calibrating each 24 hours of data, which is equivalent to averaging the total flux density over this time (Table~\ref{tab:150}). Main advantage of such an approach is that the signal-to-noise ratio can be improved not only by averaging but also by removing the residual of the extended structure and confusion. The contribution of the extended flux density repeats over each 24 hours provided the measurements are taken at similar hour angles each day and the confusing sources do not vary. 

The distribution of the differences in polarized flux density between two days in each observing session is used to estimate the actual measurement error, $\sigma \sim 2.5$ mJy, which is only weakly dependent on frequency for the sources with polarization stronger than 15 mJy. 
 
We find a number of sources with a total flux density IDV below 3$\sigma $ but with a clearly variable polarization during a given observing session (Table~\ref{tab:150}), which indicates that the variable polarized component is only a small fraction of a quiescent total flux density of such sources.

\begin{figure*}
\renewcommand{\subfigcapskip}{1pt}
\renewcommand{\subfigtopskip}{3pt}
\centering
\mbox{\subfigure[1.4 GHz]{\psfig{bbllx=0pt,bblly=0pt,bburx=675pt,bbury=500pt,file=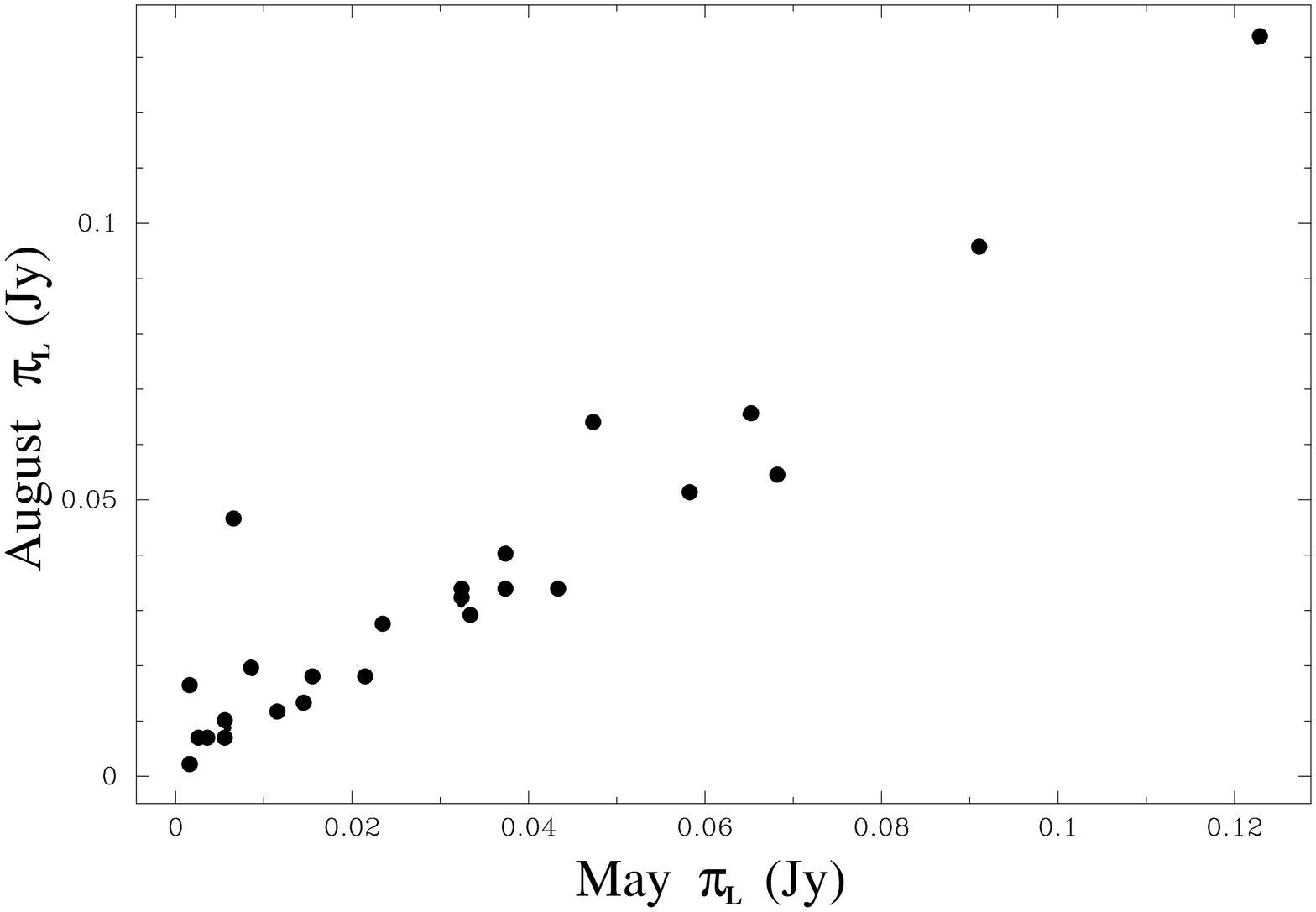,clip=,angle=0,height=60mm,width=0.48\textwidth}}\quad
\subfigure[2.4 GHz]{\psfig{bbllx=0pt,bblly=0pt,bburx=675pt,bbury=500pt,file=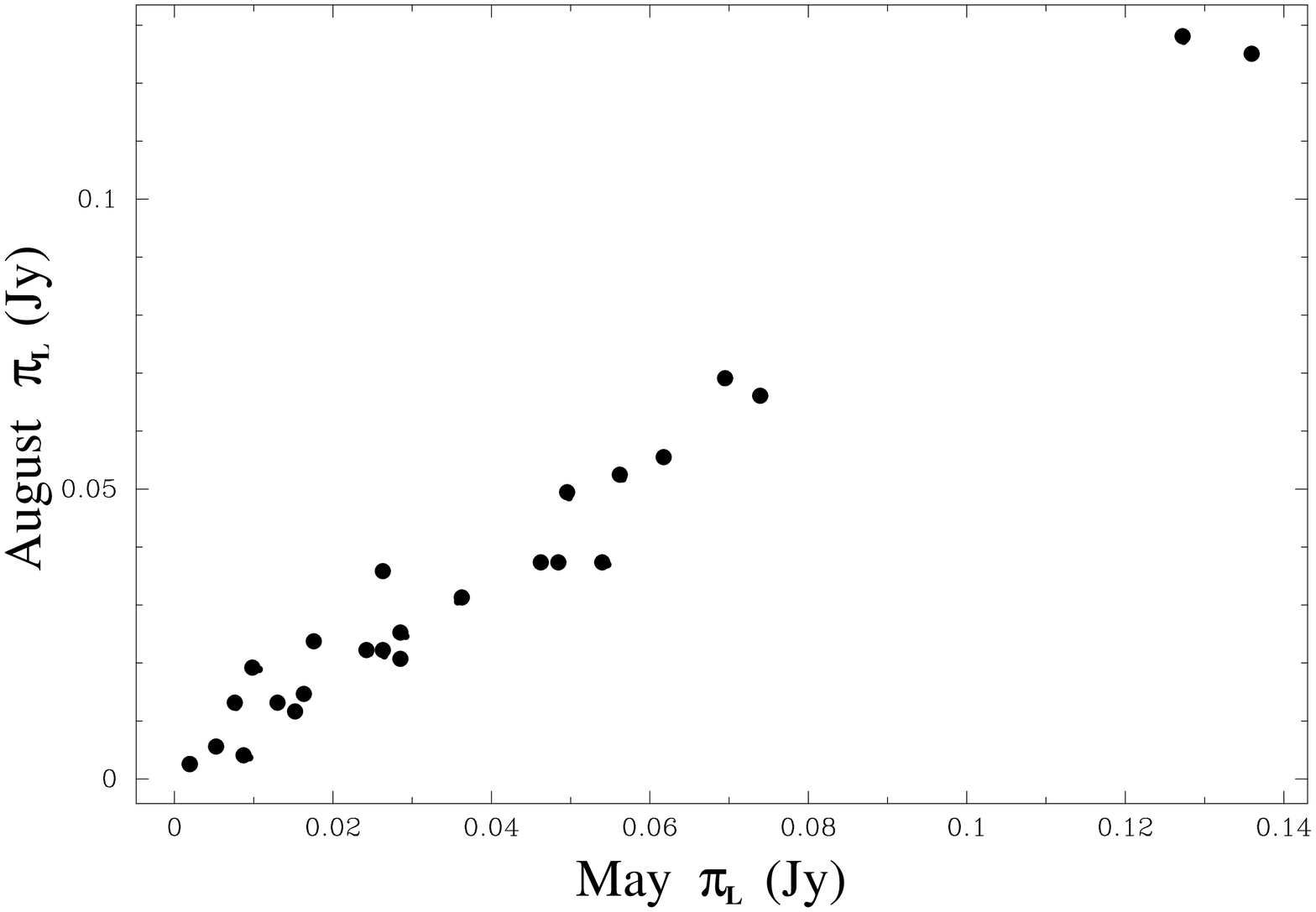,clip=,angle=0,height=60mm,width=0.48\textwidth}}}
\mbox{\subfigure[4.8 GHz]{\psfig{bbllx=0pt,bblly=0pt,bburx=675pt,bbury=500pt,file=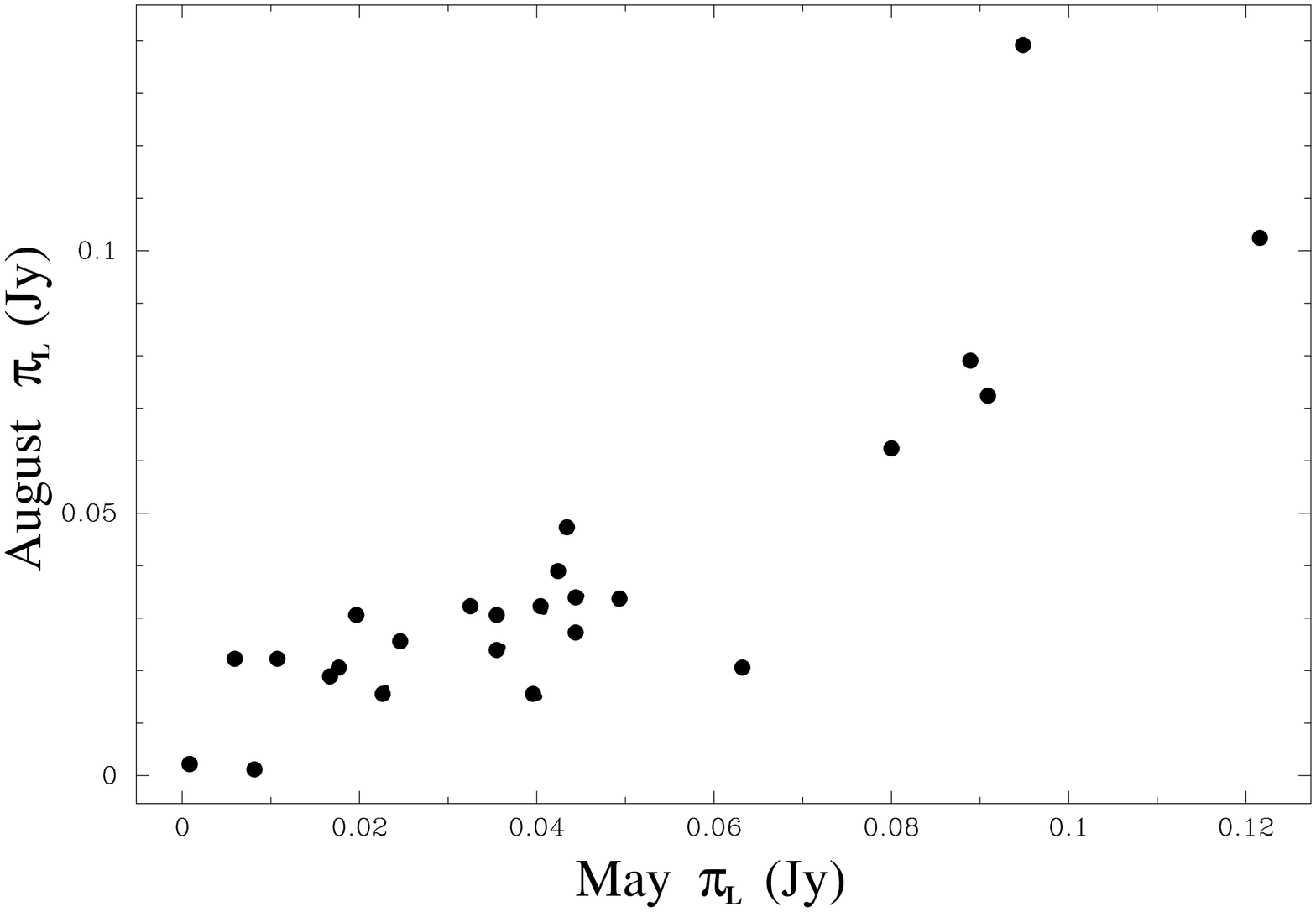,clip=,angle=0,height=60mm,width=0.48\textwidth}}\quad
\subfigure[8.6 GHz]{\psfig{bbllx=0pt,bblly=0pt,bburx=675pt,bbury=500pt,file=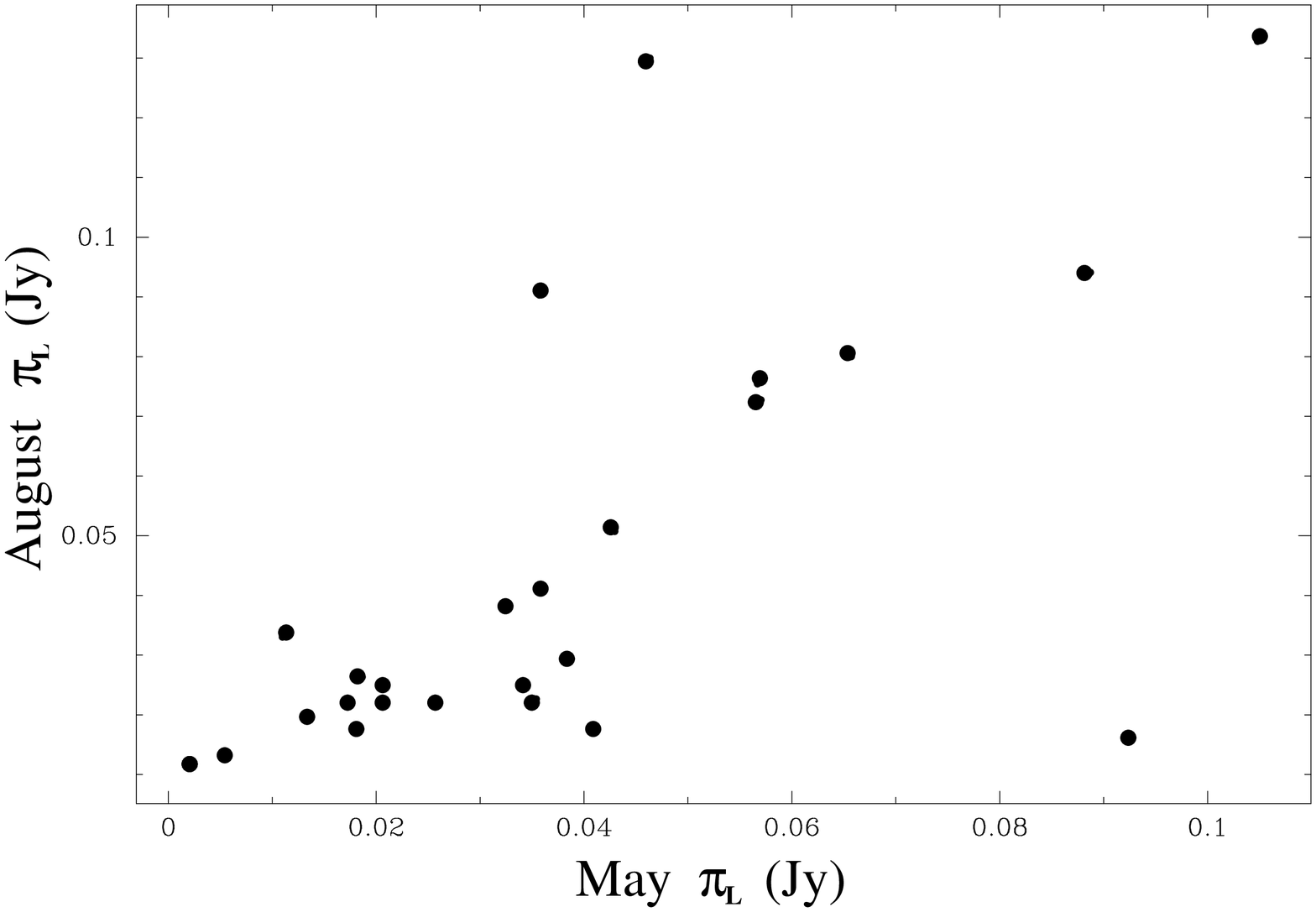,clip=,angle=0,height=60mm,width=0.48\textwidth}}}
\caption{Scatter plot of the average polarized flux density in May versus August 1994 for all frequencies.}
\label{fig102}
\end{figure*}

The changes in polarization over three months are illustrated in scatter plot (Figure~\ref{fig102}) of the averaged, polarized flux density data between May and August 1994. The linear polarization appears to be more variable at high frequencies (4.8 and 8.6 GHz) similar to the monthly variations of the total flux density. 
Note that two sources (PKS~0646-306 and PKS~1610-771) show monthly changes in the polarized flux density without a significant change in their total flux density. 
We find that only three sources, PKS~0364-279, PKS 0537-441 and PKS~0607-157, show variable polarization on both the monthly and daily timescales.

\section{Statistical Properties of IDV}
\label{stat_idv}
Various models of IDV phenomenon predict that the amplitudes and timescales of variability can be related to the structure of the radio source, its properties or its position in relation to the plane of our Galaxy. 

\subsection{Dependence of modulation index on spectral index}
Heeschen et al. (1987) showed that flat-spectrum sources are more likely to show intraday variability. We narrow that conclusion by finding that in our selection of flat-spectrum sources, those with inverted spectra over the observed range of frequencies, are more likely to show IDV than sources with other shapes of spectra. 
The flux densities of such sources are dominated by the most compact objects.  This is an additional indicator that the rapid variability originates in the smallest source components.

The modulation index as a function of the four spectral indices (Table~\ref{tab:6}) was used to search for a correlation between high modulation index and inverted spectra. 
We calculated the mean spectral index for all non-variable sources ($\mu < 3\sigma $) at a given frequency, and compared it with the mean spectral index of IDVs ($\mu \geq 3\sigma $) for $\alpha_{4.8}^{8.6}$, $\alpha_{2.4}^{4.8}$ and $\alpha_{1.4}^{8.6}$. The mean spectral index differences for these two groups of sources are -0.12, -0.06 and -0.05 respectively. The only significant difference occurred for high frequency spectral index, $\alpha_{4.8}^{8.6}$.

The lack of strong correlation between the spectral index and modulation indices suggests that the IDV strength is not a simple function of the source size.   

\subsection{Intraday variability as a function of Galactic coordinates}
The interstellar scintillation (ISS) model predicts different amplitudes and timescales of IDV depending on the source position with respect to the plane of our Galaxy. The detailed properties of variability depend on the strength of the scattering in the Galactic medium as well as a size of the scintillating source (see eg. Narayan 1992, Walker 1998). The scattering strength is related to the amount of turbulence in the Galactic medium and it is derived mostly from observations of the scattering measure in pulsars. A model of the Galactic interstellar medium was developed by Taylor and Cordes (1993) from pulsar observations. 
Applying this model to the theory of ISS (Walker 1998), the following predictions can be made:
\begin{itemize}
\item[(i)] Variability of a point source is strongest and most rapid at {\em the transition frequency}, which corresponds to a scattering strength equal to unity. The transition frequency increases towards the Galactic plane and ranges from a few GHz near the Galactic poles to a few tens of GHz at the Galactic plane.
\item[(ii)] The timescale of variability at the transition frequency is of the order of a few hours and does not depend strongly on the Galactic position. 
\item[(iii)]  The angular size of a variable source has to be comparable to the size of the first Fresnel zone, $\theta _{F}$ at the transition frequency. The predicted source sizes are of the order of $\mu$arcsecs. Sources larger than $\theta _{F}$ scintillate with decreased amplitude and longer timescale.  
\end{itemize} 

\begin{figure}
\renewcommand{\subfigcapskip}{5pt}
\renewcommand{\subfigtopskip}{0pt}
\centering
\psfig{bbllx=0pt,bblly=0pt,bburx=745pt,bbury=530pt,file=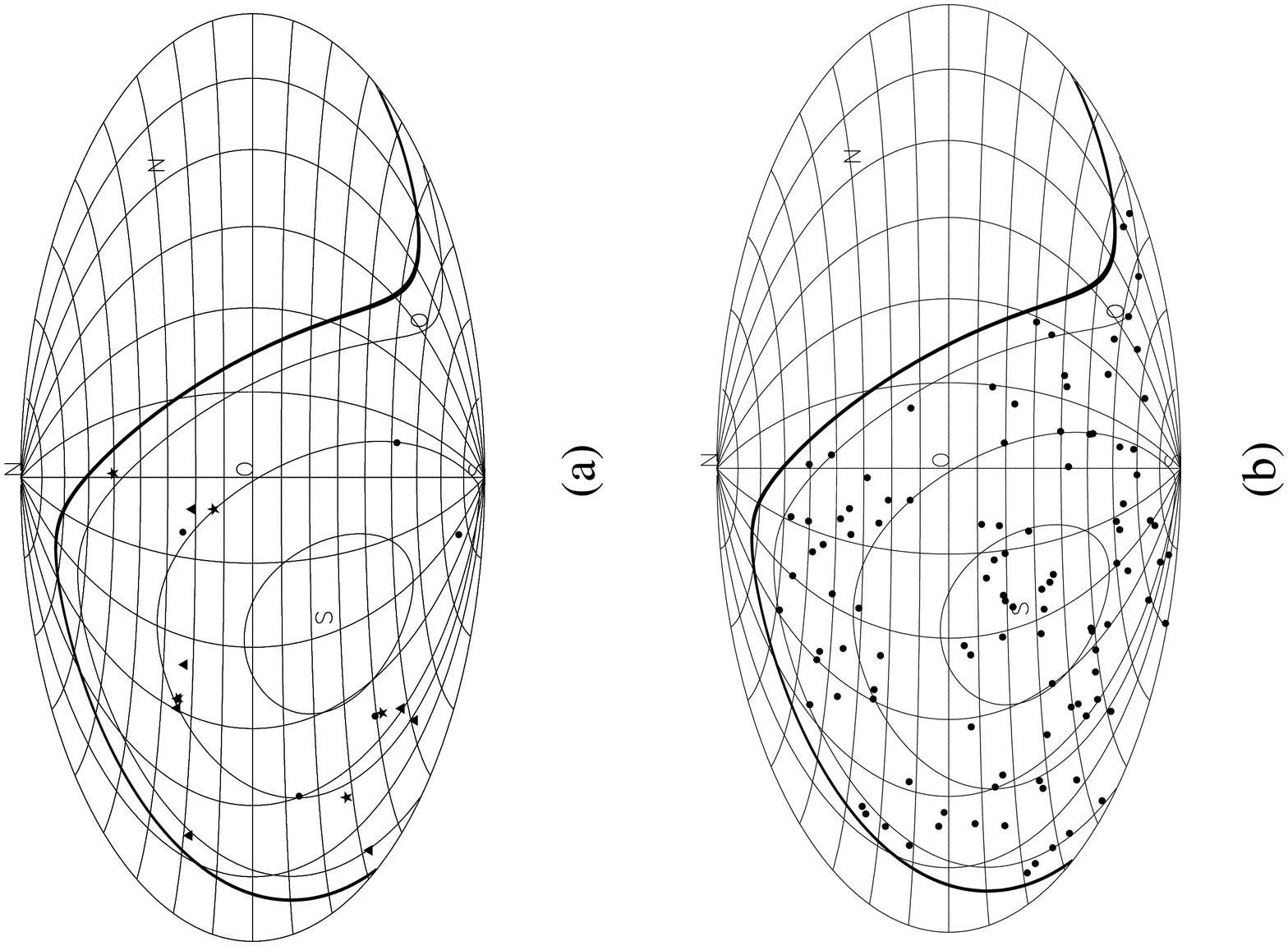,angle=270,height=120mm,width=0.48\textwidth}
\caption {(a) Distribution of IDV sources with respect to the Galactic plane. Different symbols correspond to sources which show strongest variability at different frequencies: stars (8.6 GHz), triangles (4.8 GHz) and dots (2.4 GHz). (b) Distribution of the whole sample in the Galactic coordinates. Data points are shown for May 1994 at 4.8 GHz. Only sources satisfying 'pointlikeness' criteria (see Table 1) were included. The 1.4 GHz data is not included here due to the larger uncertainties in flux density measurements. The thick curve shows the limits of the Survey, $\delta =+10^{\circ }$.}
\label{fig29}
\end{figure}

The sample of IDV sources found during the ATCA IDV Survey provides an opportunity to test the above model predictions. The Galactic positions of IDV sources from table~\ref{tab:11} at one of the three frequencies 8.6, 4.8 and 2.4 GHz (Figure~\ref{fig29}(a)) are compared with the Galactic distribution of all sources in the Survey (Figure~\ref{fig29}(b)). 

The ISS model predicts that the maximum variability for a source with $\theta \leq \theta _{F}$ should occur at the transition frequency, which in turn is a strong function of the Galactic latitude. From the Figure~\ref{fig29}(a) it is quite clear that the distribution of variability maxima, which are marked with different symbols for different frequencies, does not follow this prediction. The maxima of variability at different frequencies appear randomly distributed in otherwise non-random region where IDV prevails.
day-night effect (Kedziora-Chudczer 1998). 

The reasons which may contribute to the lack of clear dependence of the frequency of strongest IDV on the Galactic latitude are the deficiencies of the ISM model, the source size effects and the uncertainties in measurement of modulation indices. 
The Taylor \& Cordes model is most accurate at latitudes close to Galactic plane, where the number of pulsars is largest. An excess of IDV sources in the swathe between $210^{\circ}$ and $250^{\circ}$ of Galactic longitude reaching towards high Galactic latitudes may correspond to real features of the ISM which are not described well by the pulsar-based model. 
 
The modulation index and timescale of variability depends on the size of the source. In figure~\ref{fig29} the maximum modulation index is assumed to occur at the transition frequency. However if the intrinsic size of the radio source increases with decreasing frequency as $\theta _{S}\propto\nu^{-1}$ (Kellermann \& Owen 1988), the peak of variability may occur at lower frequency than expected for the point source at a given latitude.
This may explain the presence of low frequency '$\mu$-peakers' at low Galactic latitudes. The sources, which show the highest $\mu$ at latitudes higher than expected from the Taylor \& Cordes ISM model, may be explained, if their compact, variable components become self-absorbed at low frequencies. 
   
In addition, the timescale of 96 hours may not be sufficient to sample over many scintles, especially at lower frequencies. Therefore our measured modulation indices are in many cases only lower limits of the typical modulation indices of the scintillating sources.

\section{Discussion}
\subsection{Properties of IDV sources}  
The properties of IDV observed in any Survey may depend on the choice of the parent sample. 
Although the flat-spectrum flux density-limited sample of sources used in our Survey was selected on the basis their radio compactness, without distinction based on the type of the object, the majority of the selected sources are blazars (ie. flat spectrum radio quasars (FSRQ) and radio selected BL Lac objects), which tend to be both, compact and luminous (Urry \& Padovani 1995). In order to show IDV  the sources have to be compact  regardless which interpretation of IDV is preferred. The size of the source is dictated
by causality arguments in case of intrinsic variability. Similarly the ISS effects are strongest for the very compact sources. 

The IDV sources found in the ATCA Survey consist of 6 BL Lacs and 16 quasars. This is almost 50\% of the BL Lacs from the original sample, as compared with only $20\%$ of quasars. Notably the Survey sample contains 7 radio sources identified as galaxies, none of which showed intraday variability.

Modulation indices derived for the IDV sources are typically less than 10\%. We found that only a small fraction of the sources (2.5\%) show occasional stronger variations. Variability appears to be quasiperiodic, which is evident from well sampled lightcurves, such as for PKS~0405$-$385 (Kedziora-Chudczer et al. 1997). 
The largest modulation indices are typically found at 4.8 and 2.4 GHz.  

The present Survey was designed to explore timescales between 2 hours and $\sim3$ days.  Characteristic timescales were found to depend on the observed frequency. For best sampled lightcures of a few IDV sources in our sample it is evident that the timescale of variability decreases with frequency.
 
In present survey we find that most of the IDV sources are at relatively low redshifts. 
If we interpret IDV as a propagation effect, it should be more frequent for the more distant sources of the same size. For instance, ISS depends on the distance of the source only through its angular size. However the compact, scintillating components of a distant source have to be sufficiently bright in order to measure any flux density changes. The very compact components of the nearby sources can satisfy such a signal-to-noise requirement more easily.

The following properties of IDV sources were identified at different wavelengths.
A substantial number (40\%) of our IDV sources are known to be optically variable objects, which show high and variable optical polarization.  
Four of them (PKS~0422$+$004, PKS~0808$+$019, PKS~1144$-$379 and PKS~2155$-$304) are BL Lacs, the others (PKS~0440$-$003, PKS~0528$-$250, PKS~0607$-$157, PKS~0646$-$306 and PKS~1622$-$297) are quasars. Most of the IDVs in our sample are only weak sources of X-ray and $\gamma$-ray emission, which is a selection bias rather than a real effect. One of the more unusual sources in this respect is PKS~1622$-$297, which showed a strong $\gamma$-ray outburst when its energy increased $\sim200$ times within a few days (Mattox et al. 1997). 

PKS~2155$-$304 is an IDV source, which shows unusually strong X-ray emission. It is known to show rapid X-ray, optical and UV variability (Allen et al. 1993). This is the only source which was included in our Survey only for its interesting properties (it violates the flux density limit in our selection criteria). Our IDV observations of PKS~2155$-$304 were used as a part of its multiwavelength monitoring campaign (Pesce et al. 1997). An important result of this monitoring is the lack of correlation between radio and higher energy flux density changes for this source.

However the presence of correlation between optical and radio variability is claimed for another IDV source, S5~0716+714 (Quirrenbach et al. 1991). To resolve the question of correlated or otherwise variability of IDV sources at high and low energies, intensive multiwavelength monitoring of known IDV sources was undertaken recently.

\subsection{Intrinsic versus extrinsic IDV}
The suggested interpretations of IDV fall into two categories: processes intrinsic and extrinsic to the sources. The first possibility is difficult to accept for the extremely strong and rapid variability in sources such as PKS~0405$-$385, which imply the brightness temperature as high as $T_{var}\sim10^{21}$K (Kedziora-Chudczer et al. 1997).

The variability brightness temperature, $T_{var}$, can be reduced by postulating relativistic beaming $\delta$. In the most extreme cases one has to invoke $\delta>100$, which is difficult to justify on the theoretical (Begelman et al. 1994) and statistical grounds. The alternative idea of the intrinsic brightness temperature $T_{b}$ actually exceeding the Compton scattering limit for the duration of an IDV event is yet to be fully explored.

Interpretation of IDV in terms of propagation-based flux density changes can often avoid a problem of the high brightness temperature, as it puts less stringent limits on the size of the source. Two most often considered propagation effects are microlensing (Quirrenbach et al. 1991) and interstellar scintillations (Rickett 1990).
 
The typical timescales ($>1$ day) observed for IDV sources are consistent with microlensing in the presence of superluminal transverse velocities between the lens and the source, which could arise due to superluminal motion in the source (Gopal-Krishna and Subramanian 1991). However, to explain the observed hourly IDV over a few months due to microlensing events, one would also expect an optical depth for lensing high enough to cause gravitational multiple imaging of IDV sources with a few arcsecond separation, which has not been observed in any of these sources. 
   
Models based on interstellar scintillation generally require less extreme
brightness temperatures in IDV source because the source size is
determined from the Fresnel scale requirements rather than from the variability timescale. IDV in terms of ISS was discussed extensively by Rickett et al. (1995). Their model agreed well with the observed features of the data for the BL Lac S5~0917+624 under the assumption that the source diameter decreases with increasing frequency. 
The main complication of the ISS model was an interpretation of the polarized flux density variability, which required the two component source model with nearly orthogonal polarization angles of both components (Rickett et al. 1995). 
 
Taking account of the complex source structure and the little known structure of the scattering medium it is not surprising that it is difficult to make a very detailed models of ISS induced IDV. Improved VLBI imaging may reveal more about the very inner structure of the radio sources, and the galactic multiwavelength surveys will provide more detailed information about the ISM. The additional complexity is introduced by the presence of both extrinsic and intrinsic IDV occurring at the same time. 

We have applied ISS model to the variability of our most extreme IDV sources: PKS 0405$-$385 (Kedziora-Chudczer et al. 1997) and PKS 1519-273 (Macquart et al. 2000).
The spectrum of modulation indices and the timescales of variability are consistent with the model. However ISS interpretation fails to bring the $T_{B}$ down below the $10^{12}$K for these sources. Coherent emission processes are often thought of as an alternative interpretation of IDV because they can produce almost arbitrary high brightness temperature (Benford 1992, Lesch \& Pohl 1992). Yet the large number density of Langmuir photons required in any such process in the small volumes inferred for the IDV sources induces higher order scattering effects, which reduce the efficiency of escaping radiation (Melrose 1998). 

Nevertheless one should remain open to the possibility of non-synchrotron emission in inner regions of AGN, in view of the problems not only with explanation of inferred brightness temperatures, $T_{var}\sim10^{21}$K for the IDV sources but also in view of the growing number sources with  $T_{B}>10^{12}$K  of directly observed with the VLBI and VSOP (Linfield et al. 1989). 

\section{Conclusion}
The ATCA IDV Survey provided a new sample of IDV sources. We examined the variability dependence on the morphology, spectral features and position of these sources with respect to the Galactic plane. 

The IDV sources found in the ATCA Survey seem to be marginally less common than suggested in previous studies (see e.g. Quirrenbach et al. 1992, who claim that about 25\% sources from their sample shows variations larger than 2\%.) 
We find that IDV is also present in the polarized flux density.    

IDV appears to be transient in a large fraction of sources. 
However we find also continuously variable sources with the best examples being  PKS~1519--273, PKS~1034--293 and PKS~1144--379 (Kedziora-Chudczer et al. 1998).

The long term behaviour of IDV sources was studied by comparing the flux densities measured in two different observing sessions separated by 3 months. It appears that IDV sources do vary strongly on the timescale of months. This property can be potentially used in search for the IDV sources.   

The IDV variability observed in the ATCA Survey is not correlated with the strength of radio emission and only weakly correlated with spectral characteristics.  The distribution of the IDV sources with respect to the Galactic plane is not random but understandably does not follow the simple predictions of ISS theory. 

The ATCA IDV Survey was a systematic step towards an understanding of the IDV phenomenon. Although at present it is not possible to choose between competing interpretations of IDV on the purely statistical basis, the on-going study of individual IDV sources discovered in the Survey will help to achieve this purpose. 

\section{References}
\begin{description}
\item Allen, W.H., Bond, I.A., Budding, E., Conway, M.J., Daniel, A., Fenton, K.B., Fujii, H., Fujii, Z., Hayashida, N., et al.1993, PhRvD, vol.48, 466 
\item Aller, H.D., Aller, M.F., Latimer, G.E., Hodge, P.E. 1985, ApJ Suppl. vol.59, p.513
\item Begelman, M.C., Rees, M.J. \& Sikora, M. 1994, ApJL, 429, L57
%\item Bridle, A.H., 1989, in {\em Synthesis Imaging in Radio Astronomy}, A.S.P. %Conf. Ser., Vol 6, p.443, ed. R.A. Perley, F.R. Schwab and A. H. Bridle
\item Benford, G. 1992, ApJ, vol. 391, L59--L62
\item Duncan, R.A., White, G.L., Wark, R., Reynolds, J.E., Jauncey, D.L., Norris, R.P. \& Taaffe
L., 1993, Proceedings ASA vol.10. p.310
%\item Ghisellini, G., Padovani, P., Celotti, A. \& Maraschi L., 1993, ApJ, %vol.407, p.65--82
\item Gopal-Krishna, Subramanian, K., 1991 Nature, vol.349, p.766
\item Heeschen, D.S., 1984, AJ, vol.89, p.1111-1123
\item Heeschen, D.S., Krichbaum, T., Schalinski, C.J. \& Witzel, A. 1987, AJ, vol.94, p.1493-1507
\item Impey, C.D. \& Tapia, S., 1990, ApJ, vol.354, p.124--139
\item Jauncey, D., Batty, M., Gulkis, S. \& Savage, A., 1982, AJ, vol.87, 763
\item Jennison, R. C., 1958, MNRAS, 118, 276-284
\item Kedziora-Chudczer, L.L., Jauncey, D.L., Wieringa, M.H., Reynolds, J.E. \& Tzioumis, A.K., 1998, Conf. Proc. IAU 164, p.271
\item Kedziora-Chudczer, L.L., Jauncey, D.L., Wieringa, M.H., Walker, M.A., Nicolson, G.D., Reynolds, J.E. \& Tzioumis, A.K., 1997 ApJL, 490, L9
\item Kedziora-Chudczer 1998 {\em Phd Thesis}
\item Kellermann, K.I. \& Pauliny-Toth I.I.K., 1969 ApJL vol. 155, L71-L78
\item Kellermann, K.I. \& Owen F.N. 1988, in 'Galactic and Extragalactic Radio Astronomy' ed. G.L. Verschuur \& K.I. Kellermann, p.581
\item Kochanev, P.Y. \& Gabuzda, D.C., 1998, Proc. IAU Coll. 164, vol.144, p.273--274
\item Lesch, H. \& Pohl, M., 1992, A\&A, vol. 254, p. 29--38
\item Linfield, R. P., Levy, G. S., Ulvestad, J. S., Edwards, C. D., Dinardo, S. J., et al., 1989 ApJ, vol.336, p.1105
%\item McKay, D. 1994, in ATCA User's Guide, ed. S. Houghton \newline %({\em\small http://www.narrabri.atnf.csiro.au/www/observing/users\_guide.html})
\item Mattox, J.R., Wagner, S.J., Malkan, M., McGlynn, T.A., Schachter, J.F., Grove, J.E., Johnson, W.N. \& Kurfess, J.D., 1997b, ApJ, vol.476, p.692-703
\item Melrose, D.B., 1998, Ap\&SS. vol.264, 391
\item Narayan, R. 1992, Phil. Trans. R. Soc. Lond. vol.341, p.151-165
\item  Qian, S.J., Quirrenbach, A., Witzel, A., Krichbaum, T., Hummel, C.A., Zensus J.A., 1991 A\&A, vol.241, p.15
\item Quirrenbach, A., Witzel, A., Qian, S.J., Krichbaum, T.P., Hummel, C.A.,  \& Alberdi, A. 1989 A\&A vol.226, L1--L4
\item  Quirrenbach, A., Witzel, A., Wagner, S. Sanchez-Pons, F., Krichbaum, T.P., et al. 1991 ApJL vol.372, L71--L74
\item Quirrenbach, A., Witzel, A., Krichbaum, T.P., Hummel, C.A., Wegner, R., Schalinski, C.J., Ott, M., Alberdi, A. \& Rioja, M. 1992, A\&A, vol.258, p.279--284
\item Pesce, J.E., Urry, C.M., Maraschi, L., Treves, A., Grandi, P., Kollgaard, R.I., Pian, E., Smith, P.S., et al., 1997, ApJ, vol.486, p.770-783
\item Rickett, B.J. 1990, ARAA, 28, 561
\item Rickett, B.J., Quirrenbach, A., Wegner, R., Krichbaum, T.P. \& Witzel, A. 1995, A\&A vol.293, p.479-492
\item Sault, R., 1994, {\em personal communication}
\item Simonetti, J.H., Cordes, J.M. \& Heeschen D.S. 1985, ApJ, vol.296, p.46-59
\item Sinclair, M. W., Gough, R. G. 1991, International Proceedings of IREECON '91, p 381-384
\item Taylor, J.H. \& Cordes, J.M. 1993, ApJ, 430, 467
\item Urry, C.M. \& Padovani, P., 1995, PASP, vol.107, p.803-845
\item van der Laan, H., 1966, Nature 211, 1131
\item Wagner, S.J., Witzel, A., Krichbaum, T.P., Wegner, R.,
                Quirrenbach, A., Anton, K., Erkens, U., Khanna, R.
                 \& Zensus J.A., 1993, A\&A, vol.271, p.344-347 
\item Wagner S.J. \& Witzel, A. 1995, ARA\&A  vol.33, p.163-197
\item Wagner, S.J., Witzel, A., Heidt, J., Krichbaum, T.P., Qian, S.J.,
                Quirrenbach, A., Wegner, R., Aller, H., Aller, M., Anton, K.,
                Appenzeller, I., Eckart, A., Kraus, A., Naundorf, C., Kneer, R.,
                Stefeen, W. \& Zensus J.A. 1996, AJ vol.111, p.2187-2211
\item Walker, M.A. 1998, MNRAS vol.294, p.307-312
\item Wegner R.G. 1994, PhD thesis, University of Bonn, Germany
\item Witzel, A., Heeschen, D.S., Schalinski, C.J. \& Krichbaum T.P., 1986, Mitt. Astron. Ges. vol.65, p.239-241
\end{description}

\newpage
\newpage
\tablecaption[List of sources selected for IDV Survey and the structural information obtained from the closure phases and selfcalibration.]{\footnotesize List of sources selected for IDV Survey and the structural information obtained from the closure phases and self-calibration. The columns are labeled as follows: (1) B1950 Parkes Catalogue name, (2,3) new galactic coordinates: $l_{II}$ longitude and $b_{II}$ latitude calculated from equatorial (1950.0) coordinates, (4) optical identification and (5) redshift obtained from the NASA/IPAC Extragalactic Database (NED), (6) categorizes the sources at a given frequency: 8.6, 4.8, 2.4 \& 1.4 GHz (from left to right) according to a percentage of the flux density $S_{ext}$ observed in the ATCA IDV survey (see equation 1). P denotes a point source ($S_{ext}< 0.5\%$), C - compact source ($0.5\%\leq S_{ext}< 3.5\%$) and E - extended source ($S_{ext}\geq 3.5\%$), d indicates an axisymmetric structure and c denotes strong confusing sources in the field. 
Optical identification abbreviations are as follows: Q - quasar, G - galaxy, HPQ - highly polarized quasar, BL Lac - BL Lacertae type object, EF - empty field, IrS - IRAS source, Sy1 - Seyfert 1 radio galaxy.
(*) marks the sources which are known to be optically variable, ($^{\ddag}$) - obscured field, ($^{\S}$) - large differences in $S_{ext}$ between different observing sessions.}

\tablefirsthead{ \hline
Parkes    &   \multicolumn{2}{c}{Galactic coord.}  &  &     \multicolumn{4}{c}{}   \\ \cline{2-3} 
Source  &$l_{II}$   & $b_{II}$  & id  & z &  \multicolumn{4}{c}{Source}\\
 name &   (deg) &   (deg) &      &          & \multicolumn{4}{c}{structure}  \\ \hline\hline }

\tablehead{\multicolumn{6}{l}{\small \slshape continued from previous page}\\ \hline
Parkes    &   \multicolumn{2}{c}{Galactic coord.}  &  &     \multicolumn{4}{c}{}   \\ \cline{2-3} 
Source  &$l_{II}$   & $b_{II}$  & id  & z &  \multicolumn{4}{c}{Source}\\
 name &   (deg) &   (deg) &      &          & \multicolumn{4}{c}{structure}  \\ \hline\hline }
\tabletail{\hline \multicolumn{9}{l}{\small\slshape continued on next page}\\}
\tablelasttail{\hline }
\twocolumn
{\scriptsize
\begin{supertabular}{l@{\hspace{2mm}}r@{\hspace{2mm}}r@{\hspace{2mm}}c@{\hspace{2mm}}c@{\hspace{2mm}}l@{\hspace{1mm}}l@{\hspace{1mm}}l@{\hspace{1mm}}l@{\hspace{1mm}}}
\label{tab:1} 
0003$-$066 &  93.21 & $-$66.51 & G          & 0.347 &P &P &C &E   \\
0013$-$005 & 103.70 & $-$61.75 & Q          & 1.575 &C &P &C &C   \\
0023$-$263 &  42.26 & $-$84.10 & G          & 0.322 &Cd& Cd& P &E    \\
0044$-$846 & 303.14 & $-$32.80 & Q          &       &P &C &E & E   \\
0048$-$427 & 303.70 & $-$74.70 & Q          & 1.749 &P &P &C &E \\
& & & & & & & & \\
0056$-$572 & 301.09 & $-$60.17 & Q          &       &C &C &C &Ec  \\
0104$-$408 & 290.98 & $-$76.25 & Q          & 0.584 &P &P &C &E    \\
0122$-$003 & 140.66 & $-$61.75 & Q*         & 1.070 &Pc&C &E & E  \\
0131$-$522 & 288.25 & $-$63.98 & Q          &       &C &C &C &Ec  \\
0138$-$097 & 158.45 & $-$68.86 & BL Lac*    & 0.440 &E & E & E & E \\
& & & & & & & & \\
0142$-$278 & 218.02 & $-$78.26 & Q          & 1.153 &C &C &C &E  \\
0146+056   & 147.79 & $-$54.24 & Q          & 2.345 &P &P &C &Ec  \\
0150$-$334 & 241.26 & $-$75.56 & Q          & 0.610 &C &P &C &Cc  \\
0202$-$172 & 185.61 & $-$70.31 & Q          & 1.740 &P &P &C &Cc  \\
0208$-$512 & 276.27 & $-$61.91 & HPQ        & 1.003 &C &C &C &C  \\
& & & & & & & & \\
0214$-$522 & 276.14 & $-$60.55 & EF         &       &E & E & C &Cc   \\
0215+015   & 162.07 & $-$54.46 & BL Lac*    & 1.715 &P &P &Cc & E  \\
0220$-$349 & 239.75 & $-$69.20 & Q          &       &P &P &Ec & Ec     \\
0252$-$549 & 272.49 & $-$54.61 & Q          & 0.537 &C &C &Cc & E  \\
0302$-$623 & 280.23 & $-$48.81 & Q          &       &C &C &Cc &Cc   \\
& & & & & & & & \\
0308$-$611 & 278.14 & $-$49.06 & Q          &       &Cj&C &C &C   \\
0334$-$546 & 266.87 & $-$49.62 & Q          &       &C &C &C &Ec  \\
0336$-$019 & 187.76 & $-$42.63 & HPQ*       & 0.852 &P &Cc&C &Ec \\
0346$-$279 & 224.38 & $-$51.02 & Q          & 0.988 &C &C &Cc & E    \\
0355$-$483 & 256.14 & $-$48.60 & Q          & 1.005 &C &C &C &E   \\
& & & & & & & & \\
0405$-$385 & 241.18 & $-$47.93 & Q          & 1.285 &P &P &C &C    \\
0420$-$014 & 195.12 & $-$33.26 & HPQ*       & 0.915 &P &P &C &Ec   \\
0422+004   & 193.64 & $-$31.87 & BL Lac*    & 0.310 &C &C &Ec & E  \\
0426$-$380 & 240.63 & $-$43.79 & BL Lac     & 1.030 &C &Cc & Ec & Ec  \\
0434$-$188 & 216.21 & $-$37.97 & Q          & 2.702 &C &C &E & E     \\
& & & & & & & & \\
0437$-$454 & 250.76 & $-$41.85 & Q          &       &C &C &E & E     \\
0440$-$003 & 197.10 & $-$28.43 & Q          & 0.850 &C &C &C &E     \\
0450$-$469 & 252.68 & $-$39.57 & Q          &       &C &C &C &E     \\
0454$-$463 & 251.91 & $-$38.88 & Q          & 0.850 &E & E &E & E     \\
0454$-$810 & 293.75 & $-$31.42 & Q          & 0.444 &P &P &P &C      \\
& & & & & & & & \\
0457+024   & 197.00 & $-$23.40 & Q          & 2.384 &P &P &P &C       \\
0502+049   & 195.40 & $-$21.03 & Q          &       &C &C &Ec &E     \\
0522$-$611 & 270.14 & $-$34.19 & Q          & 1.400 &C &C &C &Ec     \\
0528$-$250 & 228.19 & $-$28.23 & Q*         & 2.779 &C &P &C &E    \\
0537$-$286 & 232.77 & $-$27.46 & Q          & 3.104 &C &C &E &E     \\
& & & & & & & & \\
0537$-$441 & 250.06 & $-$31.15 & BL Lac*    & 0.894 &P &Pc& Cc&Cc  \\
0607$-$157 & 222.57 & $-$16.28 & Q          & 0.324 &P &P &Ec&C   \\
0642$-$349 & 244.20 & $-$16.59 & Q          & 2.165 &C &C &C &Ec   \\
0646$-$306 & 240.43 & $-$14.14 & Q          & 0.455 &P &P &C &C     \\
0648$-$165 & 227.63 & $-$7.72  &EF$^{\ddag}$&       &P &P &E &E     \\
& & & & & & & & \\
0727$-$115 & 227.58 &  2.97    &EF$^{\ddag}$&       &P &C &C &C      \\
0728$-$320 & 245.70 &  $-$6.68 & Q          &       &C &P &E &E     \\
0733$-$174 & 233.45 &     1.38 &EF$^{\ddag}$&       &C &P &P &C       \\
0736+017   & 216.94 &    11.21 & HPQ*       & 0.191 &P &P &C &P      \\
0808+019   & 220.63 &    18.37 & BL Lac*    &       &Ed$^{\S}$&P &C &E    \\
& & & & & & & & \\
0829+046   & 220.73 &    24.26 & BL Lac*    & 0.180 &P &P &Cc & Ec  \\
0834$-$201 & 243.51 &    12.16 & Q          & 2.752 &P &C &C &E     \\
0837+035   & 222.85 &    25.49 & BL Lac     &       &C &C &C &E     \\
1021$-$006 & 245.05 &    44.76 & Q*         & 2.547 &C &C &Cc & Ec     \\
1034$-$293 & 270.76 &    24.74 & BL Lac     & 0.312 &P &C &E & E     \\
& & & & & & & & \\
1036$-$154 & 261.94 &    36.41 & Q          &       &C &C &E & E     \\
1048$-$313 & 274.74 &    24.60 & EF         &       &C &P &C &C      \\
1057$-$797 & 297.93 & $-$18.22 & Q          &       &P &P &Cc &C       \\
1105$-$680 & 293.47 &  $-$7.34 & Q          & 0.588 &P &C &C &E     \\
1115$-$122 & 270.00 &    44.31 & Q/IrS      &       &C &C &C &E      \\
& & & & & & & & \\
1127$-$145 & 275.07 &    43.62 & Q          & 1.187 &Ed&Pc$^{\S}$&Cc&Cc    \\
1143$-$245 & 284.44 &    35.72 & Q          & 1.950 &E &C &C &E     \\
1144$-$379 & 289.13 &    22.95 & BL Lac*    & 1.048 &P &Pc &C &E     \\
1148$-$001 & 272.44 &    58.84 & Q*         & 1.976 &P &P &C &Ec  \\
1148$-$671 & 297.10 &  $-$5.20 & Q          &       &P &C &E &C     \\
& & & & & & & & \\
1243$-$072 & 300.38 &    55.37 & Q          & 1.286 &P &C &C &C      \\
1253$-$055 & 304.83 &    57.09 & HPQ*       & 0.538 &C &P &E &E     \\
1255$-$316 & 304.49 &    30.98 & Q          & 1.924 &P &C &C &E     \\
1256$-$220 & 305.14 &    40.57 & Q          & 1.306 &C &C$^{\S}$&C &E     \\
1334$-$127 & 319.67 &    48.44 & BL Lac*    & 0.539 &P &C &C &C      \\
& & & & & & & & \\
1351$-$018 & 332.55 &    57.21 & Q          & 3.707 &C &C &C &E     \\
1354$-$152 & 325.28 &    44.58 & Q          & 1.890 &C &C &C &Ec    \\
1402$-$012 & 337.50 &    56.44 & Q*         & 2.518 &C &C &C &E    \\
1406$-$076 & 333.81 &    50.35 & Q          & 1.494 &P &C &E &E      \\
1435$-$218 & 333.19 &    34.53 & Q          & 1.194 &C &C &Cc &E     \\
& & & & & & & & \\
1443$-$162 & 338.77 &    38.38 & Q          &       &C &C &E &E     \\
1502+036   &   2.03 &    50.34 & Q          & 0.411 &C &P &C &E     \\
1504$-$166 & 343.64 &    35.17 & HPQ*       & 0.876 &P &P &C &E      \\
1519$-$273 & 339.48 &    24.51 & BL Lac     &       &P &C &C &C      \\
1535+004   &   5.94 &    41.92 & EF         &       &P &C &C &E     \\
& & & & & & & & \\
1540$-$828 & 308.27 & $-$22.02 & EF         &       &C &P &C &C      \\
1549$-$790 & 311.22 & $-$19.39 & G          & 0.15  &P &Cc&P &E      \\
1555$-$140 & 356.43 &    28.79 & GPair      & 0.097 &C &C &C &E      \\
1556$-$245 & 348.33 &    21.27 & Q          & 2.813 &C &C &E &E      \\
1610$-$771 & 313.45 & $-$18.77 & Q*         & 1.710 &P &P &Cc &Ec     \\
& & & & & & & & \\
1619$-$680 & 320.76 & $-$12.96 & Q          & 1.360 &P &C &C &Ec     \\
1622$-$297 & 348.71 &    13.50 & Q          & 0.815 &P$^{\S}$&P &Cc&C    \\
1718$-$649 & 326.98 & $-$15.73 & G          & 0.014 &P &P &Pc&C       \\
1741$-$038 &  21.56 &    13.21 & HPQ        & 1.054 &P &C &P &C     \\
1758$-$651 & 328.84 & $-$19.57 & G          &       &P &P &C &C      \\
& & & & & & & & \\
1815$-$553 & 339.22 & $-$17.68 & Q          &       &P &P &C &E      \\
1921$-$293 &   9.32 & $-$19.45 & HPQ*       & 0.352 &P &P &Pc&Pc \\
1925$-$610 & 335.85 & $-$28.02 & Q          &       &P &C &C &E     \\
1937$-$101 &  29.35 & $-$15.21 & Q          & 3.780 &C &P &C &E      \\
1958$-$179 &  24.06 & $-$23.08 & Q          & 0.650 &P &C &C &C       \\
& & & & & & & & \\
2016$-$615 & 335.38 & $-$34.15 & Q          &       &C &C &C &E      \\
2052$-$474 & 352.65 & $-$40.24 & Q          & 1.489 &Cc&Cc & Ec & Ec    \\
2058$-$297 &  15.86 & $-$39.76 & Q          & 0.698 &C &P &C &E    \\
2106$-$413 &   0.78 & $-$42.77 & Q          & 1.055 &P &C &C &C      \\
2109$-$811 & 311.62 & $-$32.24 & EF         &       &C &P &C &E     \\
& & & & & & & & \\
2121+053   &  57.89 & $-$30.12 & Q          & 1.878 &P &P &C &C      \\
2126$-$158 &  35.95 & $-$41.74 & Q          & 3.268 &P &P &C &E     \\
2128$-$123 &  40.46 & $-$40.75 & Sy1*       & 0.501 &P &C &C &C      \\
2134+004   &  55.38 & $-$35.60 & Q*         & 1.932 &P &P &P &E     \\
2142$-$758 & 315.85 & $-$36.54 & Q          & 1.139 &C &C &C &E      \\
& & & & & & & & \\
2146$-$783 & 313.24 & $-$35.19 & Q          &       &C &C &E &E       \\
2149$-$307 &  17.08 & $-$50.78 & Q          & 2.345 &C &C &C &C           \\
2155$-$304 &  17.80 & $-$52.03 & BL Lac*    & 0.116 &C &P &C &E      \\
2243$-$123 &  53.83 & $-$56.89 & Q*         & 0.630 &P &C &C &C       \\
2245$-$328 &  13.96 & $-$62.74 & Q          & 2.268 &C &C &E &E     \\
& & & & & & & & \\
2312$-$319 &  15.05 & $-$68.50 & Q          & 0.284 &C &C &C &E      \\
2320$-$035 &  77.46 & $-$58.02 & Q          & 1.411 &P &C &C &C       \\
2326$-$477 & 336.01 & $-$64.04 & Q          & 1.299 &C &C &C &Cc  \\
2329$-$384 & 354.58 & $-$70.00 & Q          & 1.195 &C &C &C &E      \\
2332$-$017 &  84.07 & $-$58.37 & Q*         & 1.185 &C &P &E &E     \\
2333$-$528 & 326.90 & $-$60.94 & Q          &       &P &C &Cc &E     \\
2345$-$167 &  65.56 & $-$71.74 & HPQ*       & 0.576 &C &P &C &C       \\
2355$-$534 & 320.33 & $-$62.15 & Q*         & 1.006 &P &P &E &C       \\ 
\hline \hline
\end{supertabular}}
\onecolumn
\vspace{20mm}
\begin{table}[h]
\caption{The normalized scatter of PKS~1934-638 flux density measurements ($\mu_{1934-638}$), 
averaged over two observing runs, May 1994 and August 1994, at each of the four ATCA frequencies.}
\begin{center}
\begin{tabular}{cccc} \hline
  frequency & $\mu _{1934-638}$ &  theoretical rms & measured rms \\  
  (GHz) &  (\%) &  \multicolumn{2}{c}{in 1 min observation (mJy)} \\ \hline  
  8.6  & 0.35 & 0.67 & 2.4\\  
  4.8  & 0.24 & 0.72 & 1.8\\  
  2.4  & 0.50 & 0.76 & 3.9 \\  
  1.4  & 0.41 & 0.60 & 3.8\\ \hline  
\end{tabular}

\label{tab:9}
\end{center}
\end{table}
\newpage
\tablecaption{The averaged total flux densities ($\overline{S}$) and the modulation indices ($\mu_{\overline{S}}$) for the sample of IDV sources. 
N - is the number of data points included in averages. The subscripts (1) and (2) correspond to measurements for two different observing runs: May and August 1994.
The table shows the data for four frequencies: 8.6 GHz, 4.8 GHz, 2.4 GHz and 1.4 GHz.}
\tablefirsthead{ \hline
  & \multicolumn{6}{c|}{8.6 GHz} & \multicolumn{6}{c|}{4.8 GHz} & \multicolumn{6}{c|}{2.4 GHz} & \multicolumn{6}{c}{1.4 GHz} \\
  & \multicolumn{3}{c}{May 94} & \multicolumn{3}{c|}{August 94} & \multicolumn{3}{c}{May 94} & \multicolumn{3}{c|}{August 94} & \multicolumn{3}{c}{May 94} & \multicolumn{3}{c|}{August 94} & \multicolumn{3}{c}{May 94} & \multicolumn{3}{c}{August 94} \\
   Source & 
    N &
    $\overline{S}_{1}$ &  
    $\mu_{\overline{S}_{1}}$ & 
    N &
    $\overline{S}_{2}$ & 
    $\mu_{\overline{S}_{2}}$ & 
    N &
    $\overline{S}_{1}$ &  
    $\mu_{\overline{S}_{1}}$ & 
    N &
    $\overline{S}_{2}$ & 
    $\mu_{\overline{S}_{2}}$ & 
    N &
    $\overline{S}_{1}$ &  
    $\mu_{\overline{S}_{1}}$ & 
    N &
    $\overline{S}_{2}$ & 
    $\mu_{\overline{S}_{2}}$ & 
    N &
    $\overline{S}_{1}$ &  
    $\mu_{\overline{S}_{1}}$ & 
    N &
    $\overline{S}_{2}$ & 
    $\mu_{\overline{S}_{2}}$ \\
    name & & Jy &  & & Jy &    & & Jy &   & & Jy &    & & Jy &   & & Jy &   & & Jy &   & & Jy    & \\
    \hline}
  \tablehead{\multicolumn{25}{l}{\small \slshape continued from previous page}\\ \hline 
  & \multicolumn{6}{c|}{8.6 GHz} & \multicolumn{6}{c|}{4.8 GHz} & \multicolumn{6}{c|}{2.4 GHz} & \multicolumn{6}{c}{1.4 GHz} \\
  & \multicolumn{3}{c}{May 94} & \multicolumn{3}{c|}{August 94} & \multicolumn{3}{c}{May 94} & \multicolumn{3}{c|}{August 94} & \multicolumn{3}{c}{May 94} & \multicolumn{3}{c|}{August 94} & \multicolumn{3}{c}{May 94} & \multicolumn{3}{c}{August 94} \\
   Source & 
    N &
    $\overline{S}_{1}$ &  
    $\mu_{\overline{S}_{1}}$ & 
    N &
    $\overline{S}_{2}$ & 
    $\mu_{\overline{S}_{2}}$ & 
    N &
    $\overline{S}_{1}$ &  
    $\mu_{\overline{S}_{1}}$ & 
    N &
    $\overline{S}_{2}$ & 
    $\mu_{\overline{S}_{2}}$ & 
    N &
    $\overline{S}_{1}$ &  
    $\mu_{\overline{S}_{1}}$ & 
    N &
    $\overline{S}_{2}$ & 
    $\mu_{\overline{S}_{2}}$ & 
    N &
    $\overline{S}_{1}$ &  
    $\mu_{\overline{S}_{1}}$ & 
    N &
    $\overline{S}_{2}$ & 
    $\mu_{\overline{S}_{2}}$ \\
    name & & Jy & &  & Jy &   & & Jy &  & & Jy &   & & Jy &  & & Jy &   & & Jy &  & & Jy &   \\
   \hline }
\tabletail{\hline \multicolumn{25}{l}{\small\slshape continued on next page}\\}
\tablelasttail{\hline } 

{\scriptsize
\begin{supertabular}{c@{\hspace{1.5mm}}r@{\hspace{1.5mm}}r@{\hspace{1.5mm}}r@{\hspace{1.5mm}}%
r@{\hspace{1.5mm}}r@{\hspace{1.5mm}}r@{\hspace{1.5mm}}r@{\hspace{1.5mm}}r@{\hspace{1.5mm}}%
r@{\hspace{1.5mm}}r@{\hspace{1.5mm}}r@{\hspace{1.5mm}}r@{\hspace{1.5mm}}r@{\hspace{1.5mm}}r@{\hspace{1.5mm}}r@{\hspace{1.5mm}}r@{\hspace{1.5mm}}r@{\hspace{1.5mm}}%
r@{\hspace{1.5mm}}r@{\hspace{1.5mm}}r@{\hspace{1.5mm}}r@{\hspace{1.5mm}}r@{\hspace{1.5mm}}%
r@{\hspace{1.5mm}}r@{\hspace{1.5mm}}r@{\hspace{1.5mm}}r@{\hspace{1.5mm}}r@{\hspace{1.5mm}}r@{\hspace{1.5mm}}r@{\hspace{1.5mm}}r@{\hspace{1.5mm}}r@{\hspace{1.5mm}}r@{\hspace{1.5mm}}} 
\label{tab:4}
0003$-$066 & 16 & 3.06 & 0.006 & 7 & 2.85 & 0.003 & 16 & 3.17 & 0.005 & 7 & 3.08 & 0.004 &  &  &  &  &  &  &  &  &  &  &  &  & \\
0013$-$005 & 9 & 0.83 & 0.009 & 7 & 0.84 & 0.004 & 9 & 1.01 & 0.006 & 7 & 1.08 & 0.003 & 13 & 1.11 & 0.008 & 7 & 1.16 & 0.013 & 13 & 1.06 & 0.009 & 7 & 1.09 & 0.007\\
0044$-$846 & 15 & 0.50 & 0.008 &  &  &  & 15 & 0.013 & 0.035 &  &  &  &  &  &  &  &  &  &  &  &  &  &  & \\
0048$-$427 & 16 & 1.06 & 0.007 & 8 & 1.07 & 0.004 & 17 & 0.92 & 0.006 & 8 & 0.99 & 0.005 &  &  &  &  &  &  &  &  &  &  &  &  & \\
0056$-$572 & 13 & 0.52 & 0.006 & 12 & 0.50 & 0.008 & 13 & 0.58 & 0.008 & 12 & 0.57 & 0.007 &  &  &  &  &  &  &  &  &  &  &  &  & \\
 &&&&&&&&&&&&&&&&&&&&&&&&\\
0104$-$408 & 17 & 2.60 & 0.011 & 9 & 3.03 & 0.003 & 17 & 1.55 & 0.013 & 9 & 1.80 & 0.004 & 13 & 0.91 & 0.044 & 9 & 0.93 & 0.007 &  &  &  &  &  &  & \\
0131$-$522 & 13 & 0.47 & 0.008 &   &  &  & 13 & 0.010 & 0.038 &  &  & & 17 & 0.38 & 0.010 &  &  &   &  &  &  &  &  & \\
0142$-$278 & 9 & 0.72 & 0.010 & 10 & 0.62 & 0.005 & 9 & 0.76 & 0.011 & 10 & 0.74 & 0.007 & 9 & 0.71 & 0.010 & 10 & 0.73 & 0.010 &  &  &  &  &  &  & \\
0146+056 & 9 & 1.25 & 0.004 & 7 & 1.27 & 0.005 & 9 & 1.32 & 0.006 & 7 & 1.36 & 0.003 & 7 & 1.07 & 0.013 & 7 & 1.10 & 0.011 & 7 & 0.81 & 0.020 & 7 & 0.84 & 0.012\\
0150$-$334 & 17 & 0.89 & 0.006 & 10 & 0.82 & 0.004 & 17 & 1.05 & 0.007 & 10 & 1.01 & 0.003 & 12 & 1.11 & 0.017 & 9 & 1.10 & 0.008 & 12 & 1.12 & 0.017 & 9 & 1.13 & 0.012\\
&&&&&&&&&&&&&&&&&&&&&&&&\\
0202$-$172 & 9 & 1.45 & 0.017 & 10 & 1.42 & 0.008 & 9 & 1.42 & 0.007 & 10 & 1.47 & 0.009 & 11 & 1.23 & 0.024 & 10 & 1.26 & 0.008 & 11 & 1.18 & 0.017 & 10 & 1.19 & 0.018\\
0214$-$522 & 21 & 0.34 & 0.008 & 10 & 0.33 & 0.004 & 21 & 0.51 & 0.004 & 10 & 0.51 & 0.005 & 14 & 0.71 & 0.009 & 11 & 0.72 & 0.010 & 14 & 0.86 & 0.008 & 11 & 0.87 & 0.013\\
0215+015 & 10 & 0.71 & 0.007 & 9 & 0.84 & 0.010 & 10 & 0.56 & 0.014 & 9 & 0.68 & 0.008 & 7 & 0.55 & 0.021 & 9 & 0.56 & 0.018 &  &  &  &  &  &  & \\
0220$-$349 & 13 & 0.80 & 0.010 & 14 & 0.81 & 0.005 & 13 & 0.83 & 0.007 & 14 & 0.85 & 0.005 & 16 & 0.74 & 0.020 & 10 & 0.73 & 0.005 & 16 & 0.69 & 0.019 & 10 & 0.65 & 0.017\\
0252$-$549 & 17 & 0.94 & 0.008 &  &  &  &  &  &  &  &  &  &  &  &  &  &  &  &  &  &  &  &  &  & \\
&&&&&&&&&&&&&&&&&&&&&&&&\\
0302$-$623 & 12 & 1.89 & 0.003 & 12 & 1.86 & 0.006 & 12 & 2.09 & 0.007 & 12 & 2.08 & 0.004 & 18 & 2.13 & 0.009 & 13 & 2.10 & 0.008 & 18 & 2.16 & 0.009 & 13 & 1.93 & 0.005\\
0308$-$611 & 14 & 1.67 & 0.008 & 15 & 1.72 & 0.017 & 14 & 1.36 & 0.014 & 15 & 1.46 & 0.011 &  &  &  &  &  &  & 11 & 0.90 & 0.024 & 12 & 0.94 & 0.016 & \\
0334$-$546 & 12 & 0.39 & 0.004 & 11 & 0.37 & 0.004 & 12 & 0.52 & 0.005 & 11 & 0.50 & 0.005 & 14 & 0.49 & 0.007 & 16 & 0.48 & 0.005 & 14 & 0.38 & 0.016 & 16 & 0.37 & 0.035\\
0336$-$019 & 7 & 2.50 & 0.012 & 8 & 2.34 & 0.009 &  &  &  &  &  &  &  &  &  &  &  &  &  &  &  &  &  &  & \\
0346$-$279 & 9 & 1.25 & 0.009 & 11 & 1.08 & 0.007 & 9 & 1.22 & 0.030 & 11 & 1.01 & 0.015 & 7 & 1.05 & 0.016 & 11 & 0.90 & 0.022 & 7 & 0.96 & 0.027 & 10 & 0.81 & 0.030\\
&&&&&&&&&&&&&&&&&&&&&&&&\\
0355$-$483 & 13 & 0.35 & 0.005 & 34 & 0.35 & 0.038 & 13 & 0.46 & 0.009 & 34 & 0.46 & 0.005 & 11 & 0.40 & 0.017 & 35 & 0.41 & 0.009 &  &  &  &  &  &  & \\
0405$-$385 & 11 & 1.45 & 0.043 & 76 & 1.33 & 0.003 & 11 & 1.38 & 0.056 & 76 & 1.27 & 0.007 & 14 & 1.13 & 0.056 & 77 & 1.08 & 0.011 & 14 & 0.92 & 0.052 & 77 & 0.90 & 0.009\\
0420$-$014 & 6 & 2.27 & 0.012 & 11 & 2.10 & 0.011 & 6 & 2.45 & 0.006 & 11 & 2.28 & 0.005 & 6 & 2.37 & 0.007 & 15 & 2.50 & 0.008 & 6 & 2.02 & 0.013 & 14 & 2.69 & 0.011\\
0422+004 & 10 & 0.51 & 0.018 &  &  &  & 10 & 0.024 & 0.068 &  &  &  &  &  &  &  &  &  &  &  &  &  &  & \\
0426$-$380 & 15 & 1.44 & 0.008 & 22 & 1.62 & 0.004 &  &  &  &  &  &  &  &  &  &  &  &  &  &  &  &  &  &  & \\
&&&&&&&&&&&&&&&&&&&&&&&&\\
0434$-$188 & 8 & 0.98 & 0.010 & 13 & 0.97 & 0.007 & 8 & 1.07 & 0.013 & 13 & 1.06 & 0.007 &  &  &  &  &  &  &  &  &  &  &  &  & \\
0437$-$454 & 11 & 0.32 & 0.025 & 35 & 0.34 & 0.009 & 10 & 0.33 & 0.016 & 35 & 0.33 & 0.014 &  &  &  &  &  &  &  &  &  &  &  &  & \\
0440$-$003 &  &  &  & 11 & 0.029 & 0.059 &  &  & & 11 & 1.68 & 0.011 &  &  &  & 11 & 0.018 & 0.043 &  &  & & 10 & 1.48 & 0.015 \\
0450$-$469 & 15 & 0.34 & 0.012 & 21 & 0.42 & 0.003 & 15 & 0.41 & 0.016 & 21 & 0.48 & 0.004 & 11 & 0.46 & 0.034 & 24 & 0.49 & 0.006 &  &  &  &  &  &  & \\
0454$-$810 & 16 & 1.87 & 0.014 & 17 & 2.03 & 0.006 & 16 & 1.72 & 0.019 & 17 & 1.79 & 0.010 &  &  &  &  &  &  & 13 & 1.28 & 0.010 & 17 & 1.16 & 0.012 & \\
&&&&&&&&&&&&&&&&&&&&&&&&\\
0457+024 & 7 & 1.30 & 0.007 & 12 & 1.31 & 0.011 & 7 & 1.75 & 0.007 & 12 & 1.80 & 0.009 & 7 & 2.00 & 0.013 & 10 & 2.05 & 0.025 & 7 & 1.94 & 0.014 & 9 & 1.97 & 0.029\\
0502+049 & 6 & 0.64 & 0.005 & 9 & 0.72 & 0.027 & 6 & 0.81 & 0.018 & 9 & 0.77 & 0.011 &  &  &  &  &  &  & 10 & 0.91 & 0.024 & 8 & 0.85 & 0.021 & \\
0522$-$611 & 11 & 0.65 & 0.007 & 13 & 0.61 & 0.008 & 11 & 0.66 & 0.005 & 13 & 0.65 & 0.007 & 15 & 0.67 & 0.010 & 14 & 0.69 & 0.007 & 15 & 0.73 & 0.015 & 13 & 0.75 & 0.022\\
0528$-$250 & 8 & 0.63 & 0.020 & 14 & 0.63 & 0.009 & 8 & 0.86 & 0.014 & 14 & 0.86 & 0.005 &  &  &  &  &  &  &  &  &  &  &  &  & \\
0537$-$286 & 8 & 0.45 & 0.007 & 13 & 0.39 & 0.006 & 8 & 0.67 & 0.018 & 13 & 0.56 & 0.005 &  &  &  &  &  &  &  &  &  &  &  &  & \\
&&&&&&&&&&&&&&&&&&&&&&&&\\
0537$-$441 & 12 & 6.43 & 0.006 & 35 & 6.29 & 0.004 &  &  &  &  &  &  &  &  &  &  &  &  &  &  &  &  &  &  & \\
0607$-$157 & 10 & 4.02 & 0.012 & 11 & 5.74 & 0.005 & 10 & 3.20 & 0.013 & 12 & 4.15 & 0.023 & 8 & 2.40 & 0.013 & 11 & 2.80 & 0.031 & 8 & 2.43 & 0.009 & 10 & 2.61 & 0.008\\
0642$-$349 & 16 & 0.79 & 0.008 & 11 & 0.77 & 0.009 & 16 & 0.94 & 0.007 & 11 & 0.93 & 0.011 & 11 & 0.82 & 0.019 & 12 & 0.82 & 0.010 & 11 & 0.73 & 0.026 & 12 & 0.73 & 0.020\\
0646$-$306 & 11 & 1.08 & 0.014 & 16 & 0.99 & 0.012 & 11 & 1.10 & 0.024 & 16 & 1.05 & 0.011 & 13 & 0.82 & 0.037 & 11 & 0.91 & 0.024 & 13 & 0.76 & 0.021 & 11 & 0.76 & 0.032\\
0648$-$165 & 10 & 1.52 & 0.009 & 12 & 1.64 & 0.007 & 10 & 1.47 & 0.008 & 12 & 1.52 & 0.009 &  &  &  &  &  &  &  &  &  &  &  &  & \\
&&&&&&&&&&&&&&&&&&&&&&&&\\
0727$-$115 & 13 & 4.86 & 0.005 & 12 & 4.18 & 0.007 & 13 & 5.09 & 0.012 & 12 & 4.34 & 0.007 & 10 & 3.97 & 0.009 & 14 & 3.69 & 0.011 & 10 & 3.27 & 0.013 & 13 & 2.90 & 0.013\\
0728$-$320 & 13 & 0.23 & 0.026 & 11 & 0.22 & 0.011 & 13 & 0.31 & 0.034 & 11 & 0.33 & 0.007 &  &  &  &  &  &  &  &  &  &  &  &  & \\
0733$-$174 & 9 & 1.31 & 0.010 & 12 & 1.29 & 0.004 & 9 & 1.88 & 0.004 & 12 & 1.84 & 0.007 & 13 & 2.59 & 0.012 & 15 & 2.47 & 0.011 & 13 & 2.74 & 0.006 & 15 & 2.72 & 0.004\\
0736+017 & 9 & 1.32 & 0.012 & 13 & 1.03 & 0.019 & 9 & 1.41 & 0.013 & 13 & 1.22 & 0.012 & 8 & 1.53 & 0.013 & 10 & 1.65 & 0.022 & 8 & 1.75 & 0.010 & 10 & 2.02 & 0.010\\
0808+019 & 9 & 1.02 & 0.014 & 12 & 0.57 & 0.021 & 9 & 0.97 & 0.011 & 12 & 0.59 & 0.023 & 13 & 0.80 & 0.025 & 13 & 0.56 & 0.022 &  &  &  &  &  &  & \\
&&&&&&&&&&&&&&&&&&&&&&&&\\
0829+046 & 11 & 1.18 & 0.007 & 11 & 1.23 & 0.007 & 11 & 1.12 & 0.000 & 11 & 1.10 & 0.013 & 8 & 1.00 & 0.004 & 12 & 0.98 & 0.016 & 8 & 0.97 & 0.016 & 12 & 0.93 & 0.017\\
0834$-$201 & 10 & 2.24 & 0.013 & 12 & 2.13 & 0.010 & 10 & 1.84 & 0.004 & 12 & 1.75 & 0.005 & 14 & 1.64 & 0.011 & 13 & 1.62 & 0.012 &  &  &  &  &  &  & \\
0837+035 & 9 & 0.69 & 0.014 & 11 & 0.68 & 0.007 & 9 & 0.75 & 0.004 & 11 & 0.74 & 0.012 & 8 & 0.71 & 0.016 & 10 & 0.72 & 0.011 &  &  &  &  &  &  & \\
1021$-$006 & 9 & 0.52 & 0.009 & 7 & 0.51 & 0.003 & 9 & 0.77 & 0.008 & 7 & 0.76 & 0.005 & 12 & 1.01 & 0.007 & 8 & 1.03 & 0.008 & 12 & 0.99 & 0.015 & 6 & 1.01 & 0.017\\
1034$-$293 & 11 & 1.80 & 0.057 & 10 & 2.32 & 0.012 & 11 & 1.17 & 0.092 & 10 & 1.78 & 0.113 &  &  &  &  &  &  &  &  &  &  &  &  & \\
&&&&&&&&&&&&&&&&&&&&&&&&\\
1036$-$154 & 9 & 0.41 & 0.013 & 11 & 0.41 & 0.007 & 9 & 0.39 & 0.006 & 11 & 0.39 & 0.007 &  &  &  &  &  &  &  &  &  &  &  &  & \\
1048$-$313 & 15 & 0.73 & 0.026 & 10 & 0.72 & 0.019 & 15 & 0.87 & 0.011 & 10 & 0.86 & 0.025 & 11 & 0.91 & 0.023 & 9 & 0.94 & 0.010 & 11 & 0.85 & 0.030 & 9 & 0.93 & 0.015\\
1057$-$797 & 11 & 2.10 & 0.007 & 15 & 2.09 & 0.009 & 14 & 1.49 & 0.002 & 19 & 1.33 & 0.014 & 14 & 1.01 & 0.010 & 17 & 1.04 & 0.015 & \\
1105$-$680 & 12 & 1.04 & 0.008 & 15 & 1.01 & 0.007 & 12 & 1.20 & 0.003 & 16 & 1.22 & 0.004 & 11 & 0.92 & 0.010 & 14 & 1.00 & 0.009 &  &  &  &  &  &  & \\
1115$-$122 & 11 & 0.68 & 0.019 & 9 & 0.59 & 0.003 & 11 & 0.65 & 0.018 & 9 & 0.58 & 0.012 & 9 & 0.62 & 0.009 & 9 & 0.61 & 0.013 &  &  &  &  &  &  & \\
&&&&&&&&&&&&&&&&&&&&&&&&\\
1127$-$145 & 9 & 2.84 & 0.005 & 9 & 2.82 & 0.005 & 9 & 3.57 & 0.003 & 9 & 3.58 & 0.003 & 13 & 4.39 & 0.008 & 11 & 4.50 & 0.013 & 12 & 5.06 & 0.007 & 9 & 5.19 & 0.007\\
1143$-$245 & 17 & 1.03 & 0.008 & 10 & 1.03 & 0.006 & 17 & 1.43 & 0.003 & 10 & 1.40 & 0.004 & 10 & 1.68 & 0.010 & 11 & 1.66 & 0.013 &  &  &  &  &  &  & \\
1144$-$379 & 12 & 2.33 & 0.075 & 9 & 2.14 & 0.017 & 12 & 2.47 & 0.146 & 9 & 2.04 & 0.031 &  &  &  &  &  &  &  &  &  &  &  &  & \\
1148$-$001 & 9 & 1.16 & 0.006 & 9 & 1.14 & 0.002 & 9 & 1.61 & 0.002 & 9 & 1.58 & 0.002 & 11 & 2.14 & 0.005 & 8 & 2.19 & 0.004 & 11 & 2.63 & 0.004 & 8 & 2.64 & 0.009\\
1148$-$671 & 13 & 1.15 & 0.005 & 15 & 1.11 & 0.009 & 13 & 1.69 & 0.006 & 15 & 1.66 & 0.002 &  &  &  &  &  &  & 12 & 1.29 & 0.017 & 12 & 1.10 & 0.013 & \\
&&&&&&&&&&&&&&&&&&&&&&&&\\
1243$-$072 & 14 & 1.08 & 0.007 & 6 & 1.06 & 0.006 & 14 & 0.87 & 0.007 & 6 & 0.86 & 0.001 & 11 & 0.63 & 0.004 & 7 & 0.66 & 0.003 & 11 & 0.54 & 0.012 & 7 & 0.55 & 0.018\\
1255$-$316 & 15 & 1.47 & 0.009 & 8 & 1.47 & 0.013 & 15 & 1.55 & 0.016 & 8 & 1.56 & 0.009 &  &  &  &  &  &  & 18 & 1.12 & 0.020 & 11 & 0.98 & 0.017 & \\
1256$-$220 & 16 & 0.60 & 0.012 & 7 & 0.74 & 0.027 & 16 & 0.52 & 0.014 & 7 & 0.56 & 0.022 & 11 & 0.56 & 0.009 & 7 & 0.57 & 0.007 & 11 & 0.64 & 0.040 & 7 & 0.66 & 0.017\\
1334$-$127 & 10 & 5.27 & 0.008 & 7 & 4.88 & 0.019 &  &  &  &  &  &  &  &  &  &  &  &  &  &  &  &  &  &  & \\
1351$-$018 & 14 & 0.85 & 0.007 & 6 & 0.84 & 0.009 & 14 & 0.98 & 0.003 & 6 & 0.96 & 0.014 & 11 & 0.82 & 0.018 & 6 & 0.83 & 0.014 &  &  &  &  &  &  & \\
&&&&&&&&&&&&&&&&&&&&&&&&\\
1354$-$152 & 11 & 0.88 & 0.007 & 7 & 0.84 & 0.009 & 11 & 0.96 & 0.004 & 7 & 0.87 & 0.010 & 14 & 0.76 & 0.013 & 7 & 0.75 & 0.008 &  &  &  &  &  &  & \\
1402$-$012 & 14 & 0.23 & 0.013 & 7 & 0.23 & 0.009 & 14 & 0.34 & 0.007 & 7 & 0.33 & 0.005 & 11 & 0.43 & 0.017 & 7 & 0.43 & 0.018 & 11 & 0.50 & 0.027 & 7 & 0.48 & 0.014\\
1406$-$076 & 11 & 1.01 & 0.011 & 7 & 1.00 & 0.010 & 11 & 0.76 & 0.010 & 7 & 0.80 & 0.011 &  &  &  &  &  &  &  &  &  &  &  &  & \\
1435$-$218 & 14 & 0.57 & 0.008 & 8 & 0.56 & 0.010 & 14 & 0.63 & 0.004 & 8 & 0.58 & 0.007 &  &  &  &  &  &  &  &  &  &  &  &  & \\
1443$-$162 & 12 & 0.44 & 0.012 & 7 & 0.39 & 0.019 & 12 & 0.43 & 0.015 & 7 & 0.40 & 0.008 &  &  &  &  &  &  &  &  &  &  &  &  & \\
&&&&&&&&&&&&&&&&&&&&&&&&\\
1502+036 & 13 & 0.60 & 0.020 & 6 & 0.71 & 0.002 & 13 & 0.57 & 0.014 & 6 & 0.66 & 0.008 &  &  &  &  &  &  &  &  &  &  &  &  & \\
1504$-$166 & 12 & 2.54 & 0.016 & 7 & 2.49 & 0.004 & 12 & 2.71 & 0.019 & 7 & 2.64 & 0.008 &  &  &  &  &  &  &  &  &  &  &  &  & \\
1519$-$273 & 15 & 1.41 & 0.029 & 7 & 1.77 & 0.018 & 15 & 1.44 & 0.037 & 7 & 1.72 & 0.023 & 10 & 1.09 & 0.040 & 8 & 1.28 & 0.009 & 10 & 0.98 & 0.035 & 8 & 0.91 & 0.009\\
1535+004 & 10 & 0.43 & 0.007 & 7 & 0.42 & 0.003 & 10 & 0.56 & 0.004 & 7 & 0.55 & 0.006 & 15 & 0.64 & 0.009 & 7 & 0.63 & 0.009 & 15 & 0.68 & 0.014 & 7 & 0.65 & 0.018\\
1540$-$828 & 15 & 0.58 & 0.007 & 15 & 0.58 & 0.004 & 15 & 0.67 & 0.008 & 15 & 0.68 & 0.005 & 16 & 0.74 & 0.005 & 15 & 0.74 & 0.008 & 16 & 0.89 & 0.013 & 15 & 0.85 & 0.026\\
&&&&&&&&&&&&&&&&&&&&&&&&\\
1549$-$790 & 22 & 3.01 & 0.003 &  &  &  &  &  &  &  &  &  & 14 & 0.002 & 0.068 &  &  & & 14 & 5.47 & 0.009 &  &  &  \\
1555$-$140 & 9 & 0.41 & 0.015 & 8 & 0.41 & 0.004 & 9 & 0.59 & 0.007 & 8 & 0.58 & 0.004 & 9 & 0.65 & 0.006 & 7 & 0.66 & 0.009 &  &  &  &  &  &  & \\
1556$-$245 & 9 & 0.27 & 0.015 & 8 & 0.28 & 0.042 & 9 & 0.38 & 0.028 & 8 & 0.39 & 0.033 &  &  &  &  &  &  &  &  &  &  &  &  & \\
1610$-$771 & 16 & 3.17 & 0.005 & 13 & 3.27 & 0.004 & 16 & 3.19 & 0.005 & 13 & 3.25 & 0.006 &  &  &  &  &  &  &  &  &  &  &  &  & \\
1619$-$680 & 24 & 1.41 & 0.004 & 9 & 1.34 & 0.003 & 23 & 1.93 & 0.002 & 9 & 1.86 & 0.003 & 16 & 2.04 & 0.006 & 9 & 1.88 & 0.007 & 16 & 1.66 & 0.012 & 9 & 1.54 & 0.009\\
&&&&&&&&&&&&&&&&&&&&&&&&\\
1622$-$297 & 11 & 2.25 & 0.020 & 10 & 2.19 & 0.031 & 11 & 2.32 & 0.011 & 10 & 2.24 & 0.014 & 10 & 2.12 & 0.021 & 9 & 2.09 & 0.016 &  &  &  &  &  &  & \\
1718$-$649 & 14 & 3.60 & 0.003 & 9 & 3.42 & 0.004 & 14 & 4.42 & 0.002 & 9 & 4.33 & 0.004 & 17 & 4.14 & 0.000 & 9 & 4.10 & 0.006 & 18 & 3.47 & 0.004 & 9 & 3.41 & 0.005\\
1741$-$038 & 11 & 3.23 & 0.009 & 6 & 3.81 & 0.006 & 11 & 2.53 & 0.010 & 6 & 2.74 & 0.012 & 13 & 1.89 & 0.014 & 6 & 1.94 & 0.009 & 13 & 1.56 & 0.010 & 6 & 1.47 & 0.005\\
1758$-$651 & 20 & 0.70 & 0.007 & 8 & 0.67 & 0.004 & 20 & 0.66 & 0.019 & 8 & 0.63 & 0.005 & 14 & 0.61 & 0.008 & 9 & 0.58 & 0.011 & 14 & 0.69 & 0.009 & 9 & 0.65 & 0.026\\
1815$-$553 & 14 & 1.10 & 0.006 & 8 & 0.92 & 0.003 & 14 & 1.17 & 0.004 & 8 & 1.01 & 0.003 & 17 & 0.97 & 0.008 & 6 & 1.05 & 0.008 &  &  &  &  &  &  & \\
&&&&&&&&&&&&&&&&&&&&&&&&\\
1921$-$293 & 12 & 23.9 & 0.008 & 11 & 20.9 & 0.006 & 12 & 19.68 & 0.006 & 11 & 18.06 & 0.004 & 9 & 12.94 & 0.010 & 10 & 13.45 & 0.008 & 9 & 11.28 & 0.011 & 8 & 11.70 & 0.011\\
1925$-$610 & 8 & 0.98 & 0.010 & 7 & 0.91 & 0.006 & 12 & 0.96 & 0.011 & 7 & 0.95 & 0.004 & 21 & 0.74 & 0.015 & 7 & 0.76 & 0.012 & 21 & 0.69 & 0.013 & 7 & 0.70 & 0.011\\
1937$-$101 & 9 & 0.59 & 0.012 & 6 & 0.60 & 0.008 & 8 & 0.73 & 0.007 & 6 & 0.76 & 0.005 & 12 & 0.86 & 0.012 & 7 & 0.86 & 0.011 & 12 & 0.87 & 0.008 & 7 & 0.80 & 0.010\\
1958$-$179 & 9 & 1.17 & 0.013 & 6 & 0.66 & 0.004 & 9 & 1.42 & 0.010 & 6 & 0.72 & 0.004 & 8 & 1.67 & 0.007 & 6 & 0.88 & 0.006 &  &  &  &  &  &  & \\
2016$-$615 & 19 & 0.33 & 0.008 &  &  &  & 19 & 0.005 & 0.020 &  & &  & 12 & 0.70 & 0.012 &  &  &  &  &  &  &  &  &  & \\
&&&&&&&&&&&&&&&&&&&&&&&&\\
2058$-$297 & 8 & 0.60 & 0.009 & 8 & 0.61 & 0.005 & 8 & 0.70 & 0.007 & 8 & 0.68 & 0.004 & 9 & 0.71 & 0.005 & 8 & 0.66 & 0.004 & 9 & 0.53 & 0.005 & 8 & 0.63 & 0.016\\
2106$-$413 & 9 & 2.46 & 0.007 & 11 & 2.29 & 0.005 & 9 & 2.15 & 0.003 & 11 & 2.16 & 0.005 & 10 & 1.80 & 0.007 & 7 & 1.74 & 0.007 & 10 & 1.72 & 0.013 & 7 & 1.64 & 0.010\\
2109$-$811 &  &  &  & 15 & 0.60 & 0.006 &  &  & & 15 & 0.61 & 0.008 &  &  &   & 16 & 0.54 & 0.007 &  & &  & 8 & 0.46 & 0.021 \\
2121+053 & 11 & 1.03 & 0.008 & 5 & 0.97 & 0.014 & 11 & 1.25 & 0.010 & 5 & 1.14 & 0.010 & 9 & 1.34 & 0.013 & 5 & 1.16 & 0.007 & 9 & 1.33 & 0.012 & 5 & 1.07 & 0.015\\
2126$-$158 & 9 & 0.97 & 0.003 & 7 & 0.96 & 0.005 & 9 & 1.19 & 0.004 & 7 & 1.19 & 0.004 & 9 & 0.90 & 0.010 & 11 & 0.92 & 0.007 &  &  &  &  &  &  & \\
&&&&&&&&&&&&&&&&&&&&&&&&\\
2128$-$123 & 10 & 2.55 & 0.005 & 8 & 2.63 & 0.005 & 10 & 2.31 & 0.007 & 8 & 2.27 & 0.004 & 9 & 1.87 & 0.006 & 7 & 1.79 & 0.007 & 9 & 1.57 & 0.009 & 7 & 1.49 & 0.007\\
2134+004 &  &  &  & 7 & 0.005 & 0.016 &  & &  & 7 & 8.91 & 0.004 &  &   &  & 7 & 0.006 & 0.016 &  &  & & 7 & 3.43 & 0.003 & \\
2142$-$758 & 13 & 0.70 & 0.008 & 12 & 0.70 & 0.007 & 13 & 0.87 & 0.009 & 12 & 0.87 & 0.005 &  &  &  &  &  &  &  &  &  &  &  &  & \\
2146$-$783 & 17 & 0.82 & 0.005 & 12 & 0.79 & 0.003 & 17 & 1.10 & 0.005 & 12 & 1.08 & 0.004 &  &  &  &  &  &  &  &  &  &  &  &  & \\
2149$-$307 & 12 & 1.50 & 0.014 & 6 & 1.32 & 0.006 & 12 & 1.71 & 0.010 & 6 & 1.62 & 0.005 & 17 & 1.39 & 0.016 & 6 & 1.41 & 0.016 & 17 & 1.16 & 0.013 & 6 & 1.18 & 0.018\\
&&&&&&&&&&&&&&&&&&&&&&&&\\
2155$-$304 & 20 & 0.43 & 0.024 & 6 & 0.49 & 0.008 & 21 & 0.44 & 0.021 & 6 & 0.50 & 0.005 & 20 & 0.41 & 0.034 & 9 & 0.47 & 0.020 &  &  &  &  &  &  & \\
2243$-$123 & 14 & 2.43 & 0.005 & 7 & 2.32 & 0.003 & 14 & 2.35 & 0.003 & 7 & 2.25 & 0.005 & 11 & 1.95 & 0.016 & 7 & 1.90 & 0.008 & 11 & 1.78 & 0.014 & 7 & 1.76 & 0.012\\
2245$-$328 & 11 & 0.56 & 0.008 &  &  &  & 11 & 0.010 & 0.034 &  &  &  &  &  &  &  &  &  &  &  &  &  &  & \\
2312$-$319 & 9 & 0.44 & 0.006 & 8 & 0.42 & 0.008 & 9 & 0.58 & 0.007 & 8 & 0.56 & 0.005 & 12 & 0.74 & 0.011 & 7 & 0.74 & 0.009 &  &  &  &  &  &  & \\
2320$-$035 & 9 & 0.76 & 0.010 & 8 & 0.76 & 0.007 & 9 & 0.80 & 0.005 & 7 & 0.83 & 0.010 & 11 & 0.81 & 0.010 & 9 & 0.83 & 0.007 & 11 & 0.85 & 0.009 & 9 & 0.84 & 0.011\\
&&&&&&&&&&&&&&&&&&&&&&&&\\
2326$-$477 & 16 & 1.67 & 0.006 & 8 & 1.62 & 0.006 & 16 & 1.74 & 0.007 & 8 & 1.71 & 0.010 & 13 & 2.13 & 0.007 & 9 & 2.10 & 0.009 & 13 & 2.56 & 0.013 & 8 & 2.55 & 0.027\\
2329$-$384 & 10 & 0.48 & 0.006 &  &  &  & 10 & 0.007 & 0.021 &  & &  & 18 & 0.64 & 0.007 &  &  &  &  &  &  &  &  &  & \\
2332$-$017 & 11 & 0.55 & 0.004 & 7 & 0.52 & 0.007 & 11 & 0.64 & 0.009 & 7 & 0.62 & 0.011 &  &  &  &  &  &  &  &  &  &  &  &  & \\
2333$-$528 & 16 & 1.30 & 0.008 & 9 & 1.28 & 0.005 & 16 & 1.54 & 0.004 & 9 & 1.55 & 0.002 & 12 & 1.72 & 0.010 & 8 & 1.73 & 0.008 &  &  &  &  &  &  & \\
2345$-$167 & 11 & 2.73 & 0.009 & 7 & 3.06 & 0.002 & 11 & 3.01 & 0.007 & 7 & 3.18 & 0.005 &  &  &  &  &  &  & 13 & 2.44 & 0.007 & 7 & 2.50 & 0.013 & \\
2355$-$534 & 14 & 1.66 & 0.009 & 8 & 1.61 & 0.006 & 14 & 1.76 & 0.005 & 8 & 1.76 & 0.004 &  &  &  &  &  &  &  &  &  &  &  &  & \\
\end{supertabular} }
\newpage
\tablecaption{The average polarized flux densities ($\overline{P}$), the modulation indices of linear polarization ($\mu_{\overline{P}}$) and the average position angle of the polarized flux ($\psi_{2}$). The subscripts (1) and (2) correspond to measurements for May and August 1994 respectively at 8.6, 4.8, 2.4 and 1.4 GHz.}
\begin{landscape}
  \tablefirsthead{ \hline
  & \multicolumn{8}{c|}{8.6 GHz} & \multicolumn{8}{c|}{4.8 GHz} & \multicolumn{8}{c|}{2.4 GHz} & \multicolumn{8}{c}{1.4 GHz}  \\
  & \multicolumn{4}{c}{May 94} & \multicolumn{4}{c|}{August 94} & \multicolumn{4}{c}{May 94} & \multicolumn{4}{c|}{August 94} & \multicolumn{4}{c}{May 94} & \multicolumn{4}{c|}{August 94} & \multicolumn{4}{c}{May 94} & \multicolumn{4}{c}{August 94}  \\
   Source & 
     N &
    $\overline{P}_{1}$ &  
    $\mu_{\overline{P}_{1}}$ & 
    $\psi_{1}$ &
    N &
    $\overline{P}_{2}$ & 
    $\mu_{\overline{P}_{2}}$ & 
    $\psi_{2}$ &
    N &
    $\overline{P}_{1}$ &  
    $\mu_{\overline{P}_{1}}$ & 
    $\psi_{1}$ &
    N &
    $\overline{P}_{2}$ & 
    $\mu_{\overline{P}_{2}}$ & 
    $\psi_{2}$ &
    N &
    $\overline{P}_{1}$ &  
    $\mu_{\overline{P}_{1}}$ & 
    $\psi_{1}$ &
    N &
    $\overline{P}_{2}$ & 
    $\mu_{\overline{P}_{2}}$ & 
    $\psi_{2}$ &
    N &
    $\overline{P}_{1}$ &  
    $\mu_{\overline{P}_{1}}$ & 
    $\psi_{1}$ &
    N &
    $\overline{P}_{2}$ & 
    $\mu_{\overline{P}_{2}}$ & 
    $\psi_{2}$  \\
    name & & mJy & & deg & & mJy &  &deg  & & mJy & &deg & & mJy &  & deg& & mJy & &deg & & mJy &  & deg & & mJy & & deg& & mJy &  & deg  \\
    \hline}
  \tablehead{\multicolumn{33}{l}{\small \slshape continued from previous page} \\ \hline 
  & \multicolumn{8}{c|}{8.6 GHz} & \multicolumn{8}{c|}{4.8 GHz} & \multicolumn{8}{c|}{2.4 GHz} & \multicolumn{8}{c}{1.4 GHz}  \\
  & \multicolumn{4}{c}{May 94} & \multicolumn{4}{c|}{August 94} & \multicolumn{4}{c}{May 94} & \multicolumn{4}{c|}{August 94} & \multicolumn{4}{c}{May 94} & \multicolumn{4}{c|}{August 94} & \multicolumn{4}{c}{May 94} & \multicolumn{4}{c}{August 94}  \\
   Source & 
     N &     $\overline{P}_{1}$ &  
    $\mu_{\overline{P}_{1}}$ & 
    $\psi_{1}$ &
    N &
    $\overline{P}_{2}$ & 
    $\mu_{\overline{P}_{2}}$ & 
    $\psi_{2}$ &
    N &
    $\overline{P}_{1}$ &  
    $\mu_{\overline{P}_{1}}$ & 
    $\psi_{1}$ &
    N &
    $\overline{P}_{2}$ & 
    $\mu_{\overline{P}_{2}}$ & 
    $\psi_{2}$ &
    N &
    $\overline{P}_{1}$ &  
    $\mu_{\overline{P}_{1}}$ & 
    $\psi_{1}$ &
    N &
    $\overline{P}_{2}$ & 
    $\mu_{\overline{P}_{2}}$ & 
    $\psi_{2}$ &
    N &
    $\overline{P}_{1}$ &  
    $\mu_{\overline{P}_{1}}$ & 
    $\psi_{1}$ &
    N &
    $\overline{P}_{2}$ & 
    $\mu_{\overline{P}_{2}}$ & 
    $\psi_{2}$  \\
    name & & mJy & &deg & & mJy &  &deg  & & mJy & &deg & & mJy &  & deg& & mJy & &deg & & mJy &  &deg & & mJy & & deg& & mJy &  &deg  \\
   \hline }
\tabletail{\hline \multicolumn{33}{l}{\small\slshape continued on next page} \\}
\tablelasttail{\hline } 
{\scriptsize
\begin{supertabular}{c@{\hspace{1.2mm}}r@{\hspace{1.2mm}}r@{\hspace{1.2mm}}r@{\hspace{1.2mm}}%
r@{\hspace{1.2mm}}r@{\hspace{1.2mm}}r@{\hspace{1.2mm}}r@{\hspace{1.2mm}}r@{\hspace{1.2mm}}%
r@{\hspace{1.2mm}}r@{\hspace{1.2mm}}r@{\hspace{1.2mm}}r@{\hspace{1.2mm}}r@{\hspace{1.2mm}}r@{\hspace{1.2mm}}r@{\hspace{1.2mm}}r@{\hspace{1.2mm}}r@{\hspace{1.2mm}}%
r@{\hspace{1.2mm}}r@{\hspace{1.2mm}}r@{\hspace{1.2mm}}r@{\hspace{1.2mm}}r@{\hspace{1.2mm}}%
r@{\hspace{1.2mm}}r@{\hspace{1.2mm}}r@{\hspace{1.2mm}}r@{\hspace{1.2mm}}r@{\hspace{1.2mm}}r@{\hspace{1.2mm}}r@{\hspace{1.2mm}}r@{\hspace{1.2mm}}r@{\hspace{1.2mm}}r@{\hspace{1.2mm}}}  
\label{tab:5}
0003$-$066 & 16 & 35 & 0.171 &  34.1 &  7 & 71 & 0.042 &  24.5 & 16 & 65 & 0.081 &  47.5 &  7 & 94 & 0.055 &  41.8   &    &    &       &       &    &    &       &         &    &    &       &        &    &       &       &  \\
0044$-$846 & 15 & 18 & 0.139 &   9.9 &    &    &       &       & 15 & 28 & 0.056 &   6.3 &    &    &       &         &    &    &       &       &    &    &       &         &    &    &       &        &    &       &       &  \\
0104$-$408 & 17 & 66 & 0.098 &  -6.0 &  9 & 82 & 0.042 &   2.3 & 17 & 90 & 0.028 &  19.4 &  9 & 69 & 0.074 &  21.3   & 13 & 27 & 0.079 & -74.7 &  9 & 35 & 0.056 & -66.5   &    &    &       &        &    &       &       &  \\
0122$-$003 &    &    &       &       &    &    &       &       &  9 & 33 & 0.077 &  33.3 &  8 & 34 & 0.089 &  38.3   &    &    &       &       &    &    &       &         &    &    &       &        &    &       &       &  \\
0146+056   &  9 & 91 & 0.032 & -64.2 &  7 & 93 & 0.017 & -63.8 &  9 & 88 & 0.033 & -66.7 &  7 & 78 & 0.027 & -65.1   &  7 & 27 & 0.042 & -69.7 &  7 & 22 & 0.065 & -72.9   &    &    &       &        &    &       &       &  \\
 &&&&&&&&&&&&&&&&&&&&&&&&&&&&&&&& \\
0150$-$334 & 17 & 24 & 0.080 &  30.8 & 10 & 19 & 0.104 &  19.3 & 17 & 19 & 0.107 &  28.3 & 10 & 16 & 0.138 &  19.6   &    &    &       &       &    &    &       &         & 12 & 25 & 0.074 &  67.2 &  9 & 23 & 0.248 &  71.5  \\
0202$-$172 &  9 & 18 & 0.123 & -32.3 & 10 & 26 & 0.080 & -42.6 &  9 & 81 & 0.053 & -82.7 & 10 & 61 & 0.066 & -81.4   & 11 & 57 & 0.028 & -45.7 & 10 & 51 & 0.033 & -44.7   & 11 & 39 & 0.070 & -10.5 & 10 & 39 & 0.212 & -13.3  \\
0215+015   &    &    &       &       &    &    &       &       & 10 & 09 & 0.346 &  34.0 &  9 & 27 & 0.575 & -35.5   &    &    &       &       &    &    &       &         &    &    &       &        &    &       &       &  \\
0220$-$349 & 13 & 26 & 0.089 & -76.0 & 14 & 18 & 0.074 & -71.8 & 13 & 25 & 0.065 & -61.1 & 14 & 18 & 0.110 & -55.6   & 16 & 16 & 0.145 & -43.3 & 10 & 16 & 0.086 & -43.9   &    &    &       &        &    &       &       &  \\
0252$-$549 & 17 & 39 & 0.039 & -86.1 &    &    &       &       &    &    &       &       &    &    &       &         &    &    &       &       &    &    &       &         &    &    &       &        &    &       &       &  \\
 &&&&&&&&&&&&&&&&&&&&&&&&&&&&&&&& \\
0302$-$623 & 12 & 34 & 0.214 & -73.0 & 12 & 37 & 0.048 & -68.4 & 12 & 45 & 0.074 & -60.1 & 12 & 44 & 0.176 & -51.8   & 18 & 46 & 0.083 & -28.7 & 13 & 39 & 0.119 & -25.2   & 18 & 38 & 0.160 &  48.3 & 13 & 44 & 0.157 &  51.6  \\
0308$-$611 & 14 & 21 & 0.343 &   1.9 & 15 & 25 & 0.069 &  -3.6 & 14 & 25 & 0.146 &   5.2 & 15 & 26 & 0.227 &  10.6   &    &    &       &       &    &    &       &         & 11 & 44 & 0.080 & -49.8 & 12 & 30 & 0.144 & -52.8  \\
0336$-$019 &  7 & 18 & 0.240 & -37.1 &  8 & 22 & 0.150 & -71.1 &    &    &       &       &    &    &       &         &    &    &       &       &    &    &       &         &    &    &       &        &    &       &       &  \\
0346$-$279 &  9 & 30 & 0.085 &  19.0 & 11 & 19 & 0.089 &  19.7 &  9 & 39 & 0.068 &  29.7 & 11 & 24 & 0.105 &  26.5   &  7 & 28 & 0.153 &  41.7 & 11 & 28 & 0.068 &  44.3   &  7 & 27 & 0.115 & -82.6 & 10 & 27 & 0.353 & -78.4  \\
0405$-$385 & 11 & 29 & 0.186 & -22.8 & 76 & 22 & 0.071 & -26.3 & 11 & 47 & 0.112 & -24.5 & 76 & 35 & 0.068 & -21.8   & 14 & 15 & 0.553 & -12.8 & 77 & 22 & 0.087 &  -9.4   &    &    &       &        &    &       &       &  \\
 &&&&&&&&&&&&&&&&&&&&&&&&&&&&&&&& \\
0420$-$014 &  6 & 68 & 0.079 & -69.1 & 11 & 74 & 0.030 & -67.1 &  6 & 83 & 0.045 & -75.3 & 11 & 76 & 0.059 & -74.1   &  6 & 43 & 0.154 &  79.3 & 15 & 59 & 0.141 &  80.6   &  6 & 24 & 0.224 &  59.0 & 14 & 28 & 0.318 &  44.6  \\
0426$-$380 & 15 & 21 & 0.290 & -58.1 & 22 & 94 & 0.016 &  64.6 &    &    &       &       &    &    &       &         &    &    &       &       &    &    &       &         &    &    &       &        &    &       &       &  \\
0440$-$003 &    &    &       &       & 11 & 29 & 0.059 &  40.6 &    &    &       &       & 11 & 54 & 0.069 &  65.6   &    &    &       &       & 11 & 18 & 0.132 & -65.0   &    &    &       &        & 15 & 13 & 0.618 &  35.7  \\
0450$-$469 &    &    &       &       &    &    &       &       & 15 & 21 & 0.139 &  18.3 & 21 & 19 & 0.066 &  23.1   & 11 & 30 & 0.064 &  32.2 & 24 & 26 & 0.086 &  35.7   &    &    &       &        &    &       &       &  \\
0454$-$810 & 16 & 55 & 0.110 &   4.9 & 17 & 44 & 0.138 &  11.1 & 16 & 41 & 0.139 &  14.8 & 17 & 34 & 0.154 &  -2.6   &    &    &       &       &    &    &       &         & 13 & 17 & 0.221 &  31.5 & 17 & 39 & 0.260 &  42.8  \\
 &&&&&&&&&&&&&&&&&&&&&&&&&&&&&&&& \\
0457+024   &  7 & 29 & 0.126 &  17.3 & 12 & 33 & 0.070 &  19.6 &    &    &       &       &    &    &       &         &    &    &       &       &    &    &       &         &    &    &       &        &    &       &       &  \\
0502+049   &  6 & 36 & 0.051 & -38.8 &  9 & 40 & 0.052 & -38.7 &  6 & 56 & 0.028 & -32.7 &  9 & 52 & 0.032 & -29.8   &    &    &       &       &    &    &       &         & 10 & 25 & 0.160 &   3.4 &  8 & 32 & 0.289 &  11.1  \\
0537$-$441 & 12 & 42 & 0.501 & -53.9 & 35 & 135 & 0.034 & -82.7 &    &   &       &       &    &    &       &         &    &    &       &       &    &    &       &         &    &    &       &        &    &       &       &  \\
0607$-$157 & 10 & 37 & 0.246 &  10.8 & 11 & 93 & 0.078 & -10.9 & 10 & 17 & 0.358 & -29.4 & 12 & 26 & 0.604 & -46.3   &  8 & 58 & 0.091 &  29.7 & 11 & 58 & 0.075 &  29.7   &  8 & 48 & 0.106 & -33.3 & 10 & 57 & 0.216 & -31.6  \\
0642$-$349 & 16 & 23 & 0.083 &  24.0 & 11 & 21 & 0.060 &  25.0 & 16 & 14 & 0.190 &  -9.5 & 11 & 17 & 0.158 &  -9.2   &    &    &       &       &    &    &       &         &    &    &       &        &    &       &       &  \\
 &&&&&&&&&&&&&&&&&&&&&&&&&&&&&&&& \\
0646$-$306 & 11 & 32 & 0.038 & -31.6 & 16 & 24 & 0.103 & -33.4 & 11 & 48 & 0.034 & -24.7 & 16 & 34 & 0.104 & -23.6   & 13 & 19 & 0.067 &   8.9 & 11 & 14 & 0.149 &  16.6   &    &    &       &        &    &       &       &  \\
0727$-$115 & 13 & 44 & 0.269 & -19.5 & 12 & 56 & 0.077 &  77.5 & 13 & 66 & 0.100 & -64.7 & 12 & 25 & 0.411 &  30.7   & 10 & 50 & 0.104 &  77.1 & 14 & 69 & 0.100 & -77.6   & 10 & 59 & 0.142 & -75.7 & 13 & 75 & 0.113 & -71.0  \\
0736+017   &  9 & 55 & 0.037 &  35.4 & 13 & 72 & 0.018 &  28.9 &  9 & 79 & 0.028 &  38.5 & 13 & 81 & 0.018 &  40.6   &  8 & 100 & 0.034 &  63.5 & 10 & 103 & 0.025 &  62.7 &  8 & 119 & 0.049 & -83.0 & 10 & 129 & 0.082 & -83.8  \\
0808+019   &    &    &       &       &    &    &       &       &  9 & 19 & 0.227 &  -4.9 & 12 & 11 & 0.402 &  -8.2   & 13 & 22 & 0.247 & -11.4  & 13 & 10 & 0.449 &  11.3  &    &    &       &        &    &       &       &  \\ 
0829+046   & 11 & 25 & 0.112 &  20.0 & 11 & 20 & 0.103 &  23.1 & 11 & 18 & 0.061 & -11.0 & 11 & 14 & 0.150 & -14.8   &  8 & 54 & 0.028 &  -4.0  & 12 & 48 & 0.056 &  -1.6  &  8 & 47 & 0.060 &  25.7 & 12 & 43 & 0.197 &  32.1  \\
 &&&&&&&&&&&&&&&&&&&&&&&&&&&&&&&& \\
0834$-$201 & 10 & 15 & 0.261 &  26.8 & 12 & 16 & 0.085 & -18.1 &    &    &       &       &    &    &       &         & 14 & 20 & 0.586 &  59.7  & 13 & 14 & 0.171 &  62.5  &    &    &       &        &    &       &       &  \\
0837+035   &  9 & 22 & 0.057 &  19.6 & 11 & 23 & 0.050 &  21.3 &  9 & 38 & 0.038 &  25.5 & 11 & 33 & 0.064 &  25.6   &  8 & 21 & 0.104 &  58.6  & 10 & 18 & 0.110 &  60.0  &    &    &       &        &    &       &       &  \\
1021$-$006 &  9 & 17 & 0.118 & -32.6 &  7 & 17 & 0.045 & -31.2 &  9 & 19 & 0.126 & -20.1 &  7 & 17 & 0.100 & -20.8   &    &    &       &        &    &    &       &        &    &    &       &        &    &       &       &  \\
1034$-$293 & 11 & 23 & 0.259 &  25.2 & 10 & 51 & 0.145 &   2.8 & 11 & 37 & 0.199 &  14.5 & 10 & 32 & 0.274 &  -2.3   &    &    &       &        &    &    &       &        &    &    &       &        &    &       &       &  \\
1048$-$313 &    &    &       &       &    &    &       &       & 15 & 21 & 0.101 & -72.9 & 10 & 23 & 0.054 & -73.5   & 11 & 20 & 0.124 & -41.1  &  9 & 29 & 0.066 & -43.3  & 11 & 16 & 0.167 &  43.2 &  9 & 35 & 0.142 &  46.9  \\
 &&&&&&&&&&&&&&&&&&&&&&&&&&&&&&&& \\
1057$-$797 & 11 & 13 & 0.442 &  43.0 & 15 & 34 & 0.094 &  81.6 &    &    &       &       &    &    &       &         & 14 & 19 & 0.107 & -76.1  & 19 & 23 & 0.155 & -61.8  & 14 & 8  & 0.410 &  61.9 & 17 & 25 & 0.372 &  55.3  \\
1115$-$122 & 11 & 18 & 0.113 & -16.3 &  9 & 12 & 0.147 &  -4.7 & 11 & 26 & 0.072 &   8.5 &  9 & 22 & 0.087 &  13.3   &  9 & 41 & 0.024 & -12.9  &  9 & 34 & 0.086 & -16.4  &    &    &       &        &    &       &       &  \\
1127$-$145 &  9 & 91 & 0.086 & -17.7 &  9 & 108 & 10.09 & -17.8 & 9 & 152 & 0.056 & -20.1 &  9 & 132 & 0.097 & -20.0 & 13 & 84 & 0.069 &  20.9  & 11 & 76 & 0.076 &  21.5  & 12 & 43 & 0.057 &  81.7 &  9 & 229 & 0.143 &  80.3  \\
1144$-$379 & 12 & 50 & 0.244 &  32.4 &  9 & 86 & 0.272 &  47.7 & 12 & 75 & 0.059 &  33.8 &  9 & 62 & 0.146 &  60.8   &    &    &       &        &    &    &       &        &    &    &       &        &    &       &       &  \\
1148$-$001 &  9 & 44 & 0.046 & -52.2 &  9 & 47 & 0.013 & -54.5 &  9 & 77 & 0.019 & -49.0 &  9 & 67 & 0.049 & -5   & 11 & 117 & 0.023 & -43.3  &  8 & 104 & 0.026 & -43.0 & 11 & 143 & 0.050 & -37.4 &  8 & 136 & 0.097 & -40.1  \\
 &&&&&&&&&&&&&&&&&&&&&&&&&&&&&&&& \\
1243$-$072 & 14 & 21 & 0.124 &  27.2 &  6 & 13 & 0.068 &  16.1 & 14 & 26 & 0.109 &  20.6 &  6 & 20 & 0.174 &  18.9   & 11 & 23 & 0.076 &  36.5  &  7 & 19 & 0.094 &  34.6  & 11 & 27 & 0.096 &  66.4 &  7 & 19 & 0.705 &  65.1  \\
1255$-$316 & 15 & 46 & 0.085 & -54.5 &  8 & 44 & 0.036 & -54.2 & 15 & 48 & 0.051 & -58.9 &  8 & 41 & 0.054 & -59.7   &    &    &       &        &    &    &       &        & 18 & 66 & 0.034 & -20.3 & 11 & 54 & 0.168 & -20.1  \\
1256$-$220 & 16 & 15 & 0.146 &  48.1 &  7 & 16 & 0.084 &  69.0 &    &    &       &       &    &    &       &         & 11 & 20 & 0.121 &  16.6  &  7 & 18 & 0.047 &  17.6  & 11 & 23 & 0.110 & -30.3 &  7 & 28 & 0.501 & -30.9  \\
1334$-$127 & 10 & 119 & 0.164 &  78.6 &  7 & 163 & 0.102 & -75.4 &  &    &       &       &    &    &       &         &    &    &       &        &    &    &       &        &    &    &       &        &    &       &       &  \\
1351$-$018 & 14 & 18 & 0.168 & -40.2 &  6 & 14 & 0.097 & -45.0 &    &    &       &       &    &    &       &         &    &    &       &        &    &    &       &        &    &    &       &        &    &       &       &  \\
 &&&&&&&&&&&&&&&&&&&&&&&&&&&&&&&& \\
1406$-$076 & 11 & 25 & 0.085 & -44.6 &  7 & 18 & 0.078 & -63.6 & 11 & 18 & 0.160 & -45.2 &  7 & 15 & 0.190 & -54.6   &    &    &       &        &    &    &       &        &    &    &       &        &    &       &       &  \\
1435$-$218 & 14 & 14 & 0.158 &  -9.9 &  8 & 21 & 0.079 & -12.3 & 14 & 23 & 0.071 &  -5.0 &  8 & 24 & 0.090 &  -6.8   &    &    &       &        &    &    &       &        &    &    &       &        &    &       &       &  \\
1443$-$162 & 12 & 38 & 0.038 & -51.3 &  7 & 31 & 0.041 & -50.3 & 12 & 36 & 0.065 & -59.2 &  7 & 25 & 0.072 & -58.6   &    &    &       &        &    &    &       &        &    &    &       &        &    &       &       &  \\
1504$-$166 & 12 & 61 & 0.088 &  33.8 &  7 & 70 & 0.039 &  29.3 & 12 & 42 & 0.128 &  19.7 &  7 & 41 & 0.194 &  17.1   &    &    &       &        &    &    &       &        &    &    &       &        &    &       &       &  \\
1519$-$273 & 15 & 45 & 0.195 & -30.2 &  7 & 18 & 0.287 & -20.0 & 15 & 62 & 0.319 & -24.0 &  7 & 26 & 0.331 & -35.2   & 10 & 24 & 0.305 &  17.5  &  8 & 28 & 0.085 &  41.4  & 10 & 07 & 0.597 &  46.2 &  8 & 21 & 0.540 & -33.9  \\
 &&&&&&&&&&&&&&&&&&&&&&&&&&&&&&&& \\
1610$-$771 & 16 & 106 & 0.077 & -68.5 & 13 & 135 & 0.015 & -67.2& 16 & 122 & 0.054 & -71.7 & 13 & 104 & 0.088 & -68.1 &   &    &       &        &    &    &       &        &    &    &       &        &    &       &       &  \\
1622$-$297 & 11 & 100 & 0.045 &  16.4 & 10 & 108 & 0.042 & 12.8 & 11 & 119 & 0.015 &   9.9 & 10 & 121 & 0.030 &   9.0 & 10 & 165 & 0.033 & -26.3 & 9 & 142 & 0.020 & -25.3 &    &    &       &        &    &       &       &  \\
1718$-$649 &    &    &       &       &    &    &       &       & 14 & 12 & 0.240 & -10.7 &  9 & 17 & 0.350 & -20.1   & 17 & 26 & 0.448 &  36.2  &  9 & 11 & 0.384 & -15.9  & 18 & 11 & 0.221 &   7.9 &  9 & 20 & 0.684 &  10.9  \\
1741$-$038 & 11 & 29 & 0.252 &  69.0 &  6 & 25 & 0.263 & -76.9 & 11 & 21 & 0.168 &  52.5 &  6 & 31 & 0.131 &  60.3   & 13 & 22 & 0.146 &  66.4  &  6 & 21 & 0.075 &  79.2  & 13 & 17 & 0.275 & -73.1 &  6 & 16 & 0.354 & -38.9  \\
1815$-$553 & 14 & 42 & 0.059 &  81.5 &  8 & 18 & 0.040 &  80.1 & 14 & 40 & 0.052 & -84.1 &  8 & 15 & 0.352 & -36.6   &    &    &       &        &    &    &       &   \\        
 &&&&&&&&&&&&&&&&&&&&&&&&&&&&&&&& \\
1921$-$293 & 12 & 486 & 0.096 & -31.5 & 11 & 515 & 0.056 & -67.1 & 12 & 567 & 0.076 &  17.7 & 11 & 249 & 0.103 & 82.3 &  9 & 604 & 0.022 & 18.9 & 10 & 295 & 0.085 &  22.8 &  9 & 359 & 0.074 & -21.9 &  8 & 462 & 0.034 & -25.8  \\
1925$-$610 &  8 & 37 & 0.067 &  64.2 &  7 & 42 & 0.031 &  69.6 & 12 & 35 & 0.076 &  58.7 &  7 & 29 & 0.105 &  61.2   & 21 & 30 & 0.076 &  77.6  &  7 & 24 & 0.059 &  12.4  & 21 & 32 & 0.075 & -13.3 &  7 & 34 & 0.193 &  -7.2 \\
1937$-$101 &    &    &       &       &    &       &    &       &  8 & 19 & 0.116 & -22.6 &  6 & 18 & 0.142 & -28.5   &    &    &       &        &    &    &       &        & 12 & 15 & 0.413 &  64.3 &  7 & 14 & 0.352 &  68.1  \\
1958$-$179 &  9 & 26 & 0.085 &  84.3 &  6 & 21 & 0.053 &  79.9 &  9 & 39 & 0.114 &  69.0 &  6 & 28 & 0.059 &  78.0   &  8 & 44 & 0.060 &  32.5  &  6 & 36 & 0.035 &  43.2  &    &    &       &        &    &       &       &  \\
2058$-$297 &  8 & 16 & 0.111 &   1.2 &  8 & 16 & 0.101 & -85.9 &  8 & 20 & 0.053 &  43.2 &  8 & 18 & 0.076 &  83.0   &  9 & 19 & 0.064 & -73.1  &  8 & 18 & 0.065 & -76.0  &  9 & 18 & 0.076 &  65.4 &  8 & 20 & 0.384 &  46.9  \\
 &&&&&&&&&&&&&&&&&&&&&&&&&&&&&&&& \\
2106$-$413 &  9 & 58 & 0.100 &  66.8 & 11 & 65 & 0.026 &  68.0 &  9 & 41 & 0.067 &  44.8 & 11 & 22 & 0.218 &  51.6   & 10 & 51 & 0.045 &  69.6  &  7 & 47 & 0.057 &  72.9  & 10 & 61 & 0.072 &  81.4 &  7 & 54 & 0.197 &  35.0  \\
2121+053   & 11 & 36 & 0.070 &  59.3 &  5 & 17 & 0.170 &  76.1 & 11 & 45 & 0.073 &  63.7 &  5 & 25 & 0.190 &  58.7   &  9 & 35 & 0.133 &  65.4  &  5 & 31 & 0.050 &  51.5  &  9 & 42 & 0.107 &  68.4 &  5 & 25 & 0.433 &  55.8  \\
2128$-$123 & 10 & 11 & 0.572 &  18.1 &  8 & 18 & 0.123 & -12.6 & 10 & 21 & 0.222 & -61.9 &  8 & 24 & 0.188 & -63.6   &  9 & 17 & 0.131 & -44.4  &  7 & 14 & 0.147 & -46.4  &  9 & 32 & 0.068 & -52.9 &  7 & 32 & 0.229 & -5  \\
2134+004   &    &    &       &       &  7 & 100 & 0.082 & -88.2 &   &    &       &       &  7 & 47 & 0.272 &  40.5   &    &    &       &        &  7 & 33 & 0.244 &  33.9  &    &    &       &        &  7 & 14 & 0.784 &  15.4  \\
2142$-$758 & 13 & 15 & 0.124 &  73.0 & 12 & 20 & 0.069 &  72.4 & 13 & 16 & 0.124 &  70.6 & 12 & 18 & 0.139 &  78.4   &    &    &       &        &    &    &       &        &    &    &       &       &    &       &       &  \\
 &&&&&&&&&&&&&&&&&&&&&&&&&&&&&&&& \\
2149$-$307 & 12 & 73 & 0.025 &  44.1 &  6 & 79 & 0.030 &  41.0 & 12 & 83 & 0.049 &  57.1 &  6 & 73 & 0.019 &  55.4   & 17 & 38 & 0.077 &  81.3  &  6 & 33 & 0.079 &  83.9  & 17 & 28 & 0.166 &  65.8 &  6 & 28 & 0.242 &  63.0  \\
2155$-$304 &    &    &       &       &    &    &       &       & 21 & 18 & 0.237 & -24.5 &  6 & 16 & 0.085 & -26.3   & 20 & 17 & 0.236 & -20.6  &  9 & 14 & 0.337 & -26.9  &    &    &       &        &    &       &       &  \\
2243$-$123 & 14 & 52 & 0.080 &  21.1 &  7 & 48 & 0.048 &  15.9 & 14 & 14 & 0.229 &  36.6 &  7 & 22 & 0.277 &  24.8   & 11 & 31 & 0.099 & -15.6  &  7 & 24 & 0.180 & -17.5  & 11 & 36 & 0.105 & -39.1 &  7 & 36 & 0.204 & -41.4  \\
2312$-$319 &  9 & 14 & 0.096 & -83.8 &  8 & 17 & 0.056 & -82.7 &  9 & 24 & 0.090 & -85.1 &  8 & 23 & 0.081 & -82.6   & 12 & 31 & 0.077 & -74.1  &  7 & 28 & 0.059 & -72.3  &    &    &       &        &    &       &       &  \\
2326$-$477 & 16 & 47 & 0.114 & -23.3 &  8 & 52 & 0.025 & -31.3 & 16 & 89 & 0.032 & -3    &  8 & 81 & 0.063 & -29.1   & 13 & 128 & 0.017 & -19.1 &  9 & 111 & 0.039 & -19.2 & 13 & 137 & 0.040 & -61.8 &  8 & 126 & 0.108 &   1.0  \\
 &&&&&&&&&&&&&&&&&&&&&&&&&&&&&&&& \\
2329$-$384 & 10 & 25 & 0.077 &  71.7 &    &    &       &       & 10 & 28 & 0.051 &  87.5 &    &    &       &         &    &    &       &        &    &    &       &        &    &    &       &        &    &       &       &  \\
2332$-$017 & 11 & 21 & 0.167 & -24.9 &  7 & 17 & 0.103 & -22.4 & 11 & 32 & 0.084 & -33.5 &  7 & 23 & 0.060 & -31.9   &    &    &       &        &    &    &       &        &    &    &       &        &    &       &       &  \\
2345$-$167 & 11 & 60 & 0.117 &  16.2 &  7 & 48 & 0.031 &   3.3 & 11 & 55 & 0.132 &   9.7 &  7 & 43 & 0.204 &  13.7   &    &    &       &        &    &    &       &        & 13 & 49 & 0.165 &  -1.0 &  7 & 46 & 94 &   9.4  \\
2355$-$534 & 14 & 31 & 0.154 &  -4.2 &  8 & 25 & 0.073 & -15.4 & 14 & 48 & 0.126 & -10.6 &  8 & 49 & 0.107 & -13.4   &    &    &       &        &    &    &       &        &    &            &       &        &    &       &       &         \\

\end{supertabular} }
\end{landscape}
\newpage
\tablecaption{The average spectral indices for two observing sessions: May and August 1994.}
\twocolumn
\begin{center}
  \tablefirsthead{ \hline
 PKS  & \multicolumn{4}{c|}{ May 1994} & \multicolumn{4}{c}{ August 1994} \\
    Source & 
    $\alpha^{8.6}_{4.8}$ & 
    $\alpha^{4.8}_{2.4}$ & 
    $\alpha^{2.4}_{1.4}$ & 
    $\alpha^{8.6}_{1.4}$ & 
    $\alpha^{8.6}_{4.8}$ & 
    $\alpha^{4.8}_{2.4}$ & 
    $\alpha^{2.4}_{1.4}$ & 
    $\alpha^{8.6}_{1.4}$ \\
 name  & & & & & & & &  \\
    \hline}
  \tablehead{\multicolumn{9}{l}{\small \slshape continued from previous page}\\ \hline 
 PKS  & \multicolumn{4}{c|}{ May 1994} & \multicolumn{4}{c}{ August 1994} \\
    Source & 
    $\alpha^{8.6}_{4.8}$ & 
    $\alpha^{4.8}_{2.4}$ & 
    $\alpha^{2.4}_{1.4}$ & 
    $\alpha^{8.6}_{1.4}$ & 
    $\alpha^{8.6}_{4.8}$ & 
    $\alpha^{4.8}_{2.4}$ & 
    $\alpha^{2.4}_{1.4}$ & 
    $\alpha^{8.6}_{1.4}$ \\
 name  & & & & & & & &  \\
   \hline }
\tabletail{\hline \multicolumn{9}{l}{\small\slshape continued on next page}\\}
\tablelasttail{\hline } 

{\scriptsize
\begin{supertabular}{c@{\hspace{1mm}}r@{\hspace{1mm}}r@{\hspace{1mm}}r@{\hspace{1mm}}r@{\hspace{1mm}}r@{\hspace{1mm}}r@{\hspace{1mm}}r@{\hspace{1mm}}r@{\hspace{1mm}}} 
\label{tab:6}
0003$-$066 & -0.06 &  0.24 &  0.34 &  0.17 & -0.13 &  0.24 &  0.43 &  0.18 \\
0013$-$005 & -0.33 & -0.14 &  0.09 & -0.13 & -0.41 & -0.10 &  0.12 & -0.14 \\
0023$-$263 & -0.93 & -0.74 & -0.66 & -0.78 & -0.99 & -0.73 & -0.68 & -0.80 \\
0044$-$846 & -0.27 & -0.10 & -0.15 & -0.17 &       &       &       &       \\
0048$-$427 &  0.24 &  0.36 &  0.33 &  0.31 &  0.14 &  0.41 &  0.39 &  0.32 \\
 & & & & & & & &  \\
0056$-$572 & -0.20 &  0.01 &  0.10 & -0.03 & -0.23 & -0.01 &  0.09 & -0.05 \\
0104$-$408 &  0.88 &  0.76 &  0.46 &  0.71 &  0.89 &  0.94 &  0.30 &  0.73 \\
0122$-$003 & -0.09 &  0.05 &  0.02 &  0.00 & -0.08 &  0.03 &  0.10 &  0.01 \\
0131$-$522 &  0.24 &  0.12 & -0.39 &  0.01 &       &       &       &       \\
0138$-$097 & -0.08 &  0.16 &  0.09 &  0.06 & -0.03 &  0.16 &  0.25 &  0.12 \\
 & & & & & & & &  \\
0142$-$278 & -0.09 &  0.09 & -0.17 & -0.04 & -0.29 &  0.02 & -0.06 & -0.10 \\
0146+056   & -0.09 &  0.30 &  0.50 &  0.24 & -0.11 &  0.30 &  0.50 &  0.23 \\
0150$-$334 & -0.28 & -0.08 & -0.03 & -0.13 & -0.36 & -0.12 & -0.05 & -0.18 \\
0202$-$172 &  0.02 &  0.21 &  0.08 &  0.11 & -0.07 &  0.22 &  0.10 &  0.09 \\
0208$-$512 &  0.03 & -0.09 & -0.07 & -0.04 &       &       &       &       \\
 & & & & & & & &  \\
0214$-$522 & -0.70 & -0.48 & -0.36 & -0.52 & -0.74 & -0.49 & -0.36 & -0.53 \\
0215+015   &  0.40 &  0.04 & -0.16 &  0.09 &  0.35 &  0.28 &  0.10 &  0.25 \\
0220$-$349 & -0.07 &  0.17 &  0.12 &  0.08 & -0.09 &  0.21 &  0.20 &  0.12 \\
0252$-$549 &  0.06 &  0.09 & -0.02 &  0.05 &       &       &       &       \\
0302$-$623 & -0.17 & -0.03 & -0.03 & -0.07 & -0.19 &  0.00 &  0.13 & -0.02 \\
 & & & & & & & &  \\
0308$-$611 &  0.34 &  0.46 &  0.18 &  0.34 &  0.28 &  0.46 &  0.23 &  0.33 \\
0334$-$546 & -0.50 &  0.07 &  0.50 &  0.01 & -0.53 &  0.05 &  0.48 &  0.00 \\
0336$-$019 & -0.04 &  0.11 &  0.21 &  0.09 & -0.16 &  0.07 &  0.10 &  0.01 \\
0346$-$279 &  0.05 &  0.21 &  0.17 &  0.14 &  0.11 &  0.17 &  0.18 &  0.15 \\
0355$-$483 & -0.43 &  0.17 &  0.22 & -0.01 &       &       &       &       \\
 & & & & & & & &  \\
0405$-$385 &  0.08 &  0.29 &  0.38 &  0.25 &  0.08 &  0.24 &  0.33 &  0.22 \\
0420$-$014 & -0.13 &  0.04 &  0.30 &  0.07 & -0.13 & -0.14 & -0.13 & -0.13 \\
0422+004   &  0.18 &  0.19 & -0.05 &  0.11 &       &       &       &       \\
0426$-$380 &  0.37 &  0.56 &  0.37 &  0.44 &  0.14 &  0.42 &  0.46 &  0.34 \\
0434$-$188 & -0.15 &  0.25 &  0.61 &  0.23 & -0.15 &  0.29 &  0.64 &  0.26 \\
 & & & & & & & &  \\
0437$-$454 & -0.06 & -0.05 & -0.13 & -0.08 &  0.09 &  0.00 &  0.12 &  0.07 \\
0440$-$003 &       &       &       &       & -0.19 &  0.02 &  0.20 &  0.01 \\
0450$-$469 & -0.31 & -0.16 & -0.16 & -0.21 & -0.23 & -0.02 &  0.01 & -0.08 \\
0454$-$463 & -0.28 & -0.26 & -0.50 & -0.34 &       &       &       &       \\
0454$-$810 &  0.15 &  0.15 &  0.35 &  0.21 &  0.21 &  0.39 &  0.30 &  0.31 \\
 & & & & & & & &  \\
0457+024   & -0.50 & -0.19 &  0.06 & -0.22 & -0.55 & -0.18 &  0.07 & -0.22 \\
0502+049   & -0.40 & -0.22 &  0.09 & -0.19 & -0.12 & -0.21 &  0.08 & -0.09 \\
0522$-$611 & -0.02 & -0.02 & -0.17 & -0.06 & -0.12 & -0.08 & -0.17 & -0.12 \\
0528$-$250 & -0.52 & -0.43 &  0.05 & -0.32 & -0.54 & -0.38 &  0.14 & -0.28 \\
0537$-$286 & -0.67 & -0.42 & -0.09 & -0.40 & -0.59 & -0.42 &  0.02 & -0.35 \\
 & & & & & & & &  \\
0537$-$441 &  0.32 &  0.47 &  0.31 &  0.37 &  0.34 &  0.28 &  0.51 &  0.37 \\
0607$-$157 &  0.39 &  0.41 & -0.03 &  0.27 &  0.55 &  0.56 &  0.13 &  0.43 \\
0642$-$349 & -0.30 &  0.20 &  0.20 &  0.04 & -0.33 &  0.18 &  0.23 &  0.03 \\
0646$-$306 & -0.03 &  0.42 &  0.14 &  0.19 & -0.10 &  0.20 &  0.33 &  0.14 \\
0648$-$165 &  0.06 &  0.02 & -0.21 & -0.04 &  0.13 &  0.03 & -0.28 & -0.03 \\
 & & & & & & & &  \\
0727$-$115 & -0.08 &  0.36 &  0.36 &  0.22 & -0.06 &  0.23 &  0.44 &  0.20 \\
0728$-$320 & -0.54 & -0.20 &  0.12 & -0.21 & -0.70 & -0.44 &  0.26 & -0.31 \\
0733$-$174 & -0.62 & -0.46 & -0.11 & -0.40 & -0.60 & -0.42 & -0.18 & -0.41 \\
0736+017   & -0.12 & -0.11 & -0.24 & -0.15 & -0.30 & -0.43 & -0.37 & -0.37 \\
0808+019   &  0.09 &  0.28 &  0.17 &  0.19 & -0.06 &  0.08 &  0.00 &  0.01 \\
 & & & & & & & &  \\
0829+046   &  0.09 &  0.17 &  0.05 &  0.11 &  0.20 &  0.16 &  0.09 &  0.15 \\
0834$-$201 &  0.34 &  0.17 & -0.28 &  0.09 &  0.33 &  0.11 & -0.22 &  0.08 \\
0837+035   & -0.16 &  0.08 &  0.16 &  0.03 & -0.14 &  0.04 &  0.14 &  0.01 \\
1021$-$006 & -0.68 & -0.39 &  0.04 & -0.35 & -0.67 & -0.44 &  0.03 & -0.37 \\
1034$-$293 &  0.73 &  0.82 & -0.07 &  0.52 &  0.45 &  0.29 &  0.42 &  0.38 \\
 & & & & & & & &  \\
1036$-$154 &  0.10 &  0.03 & -0.10 &  0.01 &  0.08 &  0.00 & -0.03 &  0.02 \\
1048$-$313 & -0.30 & -0.06 &  0.12 & -0.08 & -0.30 & -0.15 &  0.04 & -0.14 \\
1057$-$797 &  0.01 &  0.48 &  0.71 &  0.40 &  0.05 &  0.60 &  0.45 &  0.38 \\
1105$-$680 & -0.24 &  0.39 &  0.93 &  0.35 & -0.33 &  0.28 &  0.91 &  0.27 \\
1115$-$122 &  0.08 &  0.08 & -0.17 &  0.00 &  0.03 & -0.07 & -0.15 & -0.06 \\
 & & & & & & & &  \\
1127$-$145 & -0.39 & -0.30 & -0.26 & -0.31 & -0.41 & -0.33 & -0.26 & -0.33 \\
1143$-$245 & -0.56 & -0.23 &  0.20 & -0.21 & -0.53 & -0.24 &  0.28 & -0.18 \\
1144$-$379 & -0.10 &  0.22 & -0.20 & -0.01 &  0.09 &  0.11 &  0.36 &  0.18 \\
1148$-$001 & -0.56 & -0.41 & -0.37 & -0.45 & -0.56 & -0.47 & -0.34 & -0.46 \\
1148$-$671 & -0.65 & -0.15 &  0.68 & -0.06 & -0.68 & -0.10 &  0.86 &  0.00 \\
 & & & & & & & &  \\
1243$-$072 &  0.37 &  0.46 &  0.27 &  0.38 &  0.36 &  0.38 &  0.33 &  0.36 \\
1253$-$055 &  0.32 &  0.29 &  0.15 &  0.26 &  0.29 &  0.26 &  0.41 &  0.31 \\
1255$-$316 & -0.09 &  0.29 &  0.23 &  0.15 & -0.10 &  0.38 &  0.37 &  0.22 \\
1256$-$220 &  0.25 & -0.11 & -0.22 & -0.03 &  0.49 & -0.04 & -0.26 &  0.06 \\
1334$-$127 &  0.29 &  0.56 &  0.44 &  0.44 &  0.33 &  0.35 &  0.52 &  0.39 \\
 & & & & & & & &  \\
1351$-$018 & -0.24 &  0.25 &  0.34 &  0.12 & -0.22 &  0.20 &  0.37 &  0.12 \\
1354$-$152 & -0.16 &  0.34 &  0.35 &  0.18 & -0.05 &  0.20 &  0.39 &  0.18 \\
1402$-$012 & -0.61 & -0.36 & -0.26 & -0.41 & -0.59 & -0.38 & -0.19 & -0.39 \\
1406$-$076 &  0.49 &  0.50 &  0.00 &  0.34 &  0.37 &  0.45 &  0.09 &  0.32 \\
1435$-$218 & -0.17 & -0.15 & -0.12 & -0.15 & -0.07 & -0.14 & -0.01 & -0.08 \\
 & & & & & & & &  \\
1443$-$162 &  0.04 &  0.12 & -0.20 &  0.00 & -0.03 &  0.06 & -0.10 & -0.02 \\
1502+036   &  0.07 &  0.10 &  0.22 &  0.13 &  0.12 &  0.64 &  0.48 &  0.42 \\
1504$-$166 & -0.11 & -0.02 & -0.05 & -0.06 & -0.10 &  0.02 & -0.04 & -0.04 \\
1519$-$273 & -0.03 &  0.40 &  0.20 &  0.20 &  0.04 &  0.43 &  0.62 &  0.36 \\
1535+004   & -0.46 & -0.20 & -0.10 & -0.25 & -0.45 & -0.21 & -0.03 & -0.23 \\
 & & & & & & & &  \\
1540$-$828 & -0.25 & -0.14 & -0.34 & -0.23 & -0.29 & -0.11 & -0.27 & -0.21 \\
1549$-$790 & -0.63 & -0.21 & -0.14 & -0.33 &       &       &       &       \\
1555$-$140 & -0.60 & -0.14 &  0.81 &  0.00 & -0.61 & -0.18 &  0.85 & -0.01 \\
1556$-$245 & -0.59 & -0.53 & -0.47 & -0.53 & -0.54 & -0.47 &  1.42 &  0.07 \\
1610$-$771 & -0.01 & -0.02 & -0.21 & -0.07 &  0.01 &  0.01 & -0.12 & -0.03 \\
 & & & & & & & &  \\
1619$-$680 & -0.53 & -0.08 &  0.38 & -0.09 & -0.56 & -0.01 &  0.36 & -0.08 \\
1622$-$297 & -0.05 &  0.12 &  0.06 &  0.05 & -0.04 &  0.10 & -0.18 & -0.03 \\
1718$-$649 & -0.35 &  0.09 &  0.32 &  0.02 & -0.40 &  0.08 &  0.34 &  0.00 \\
1741$-$038 &  0.41 &  0.42 &  0.35 &  0.40 &  0.56 &  0.49 &  0.51 &  0.52 \\
1758$-$651 &  0.10 &  0.11 & -0.24 &  0.00 &  0.10 &  0.13 & -0.22 &  0.01 \\
 & & & & & & & &  \\
1815$-$553 & -0.10 &  0.27 &  0.32 &  0.16 & -0.15 & -0.06 &  0.01 & -0.06 \\
1921$-$293 &  0.33 &  0.60 &  0.25 &  0.41 &  0.25 &  0.42 &  0.25 &  0.32 \\
1925$-$610 &  0.03 &  0.37 &  0.11 &  0.19 & -0.08 &  0.31 &  0.15 &  0.14 \\
1937$-$101 & -0.37 & -0.23 & -0.01 & -0.21 & -0.40 & -0.16 &  0.12 & -0.15 \\
1958$-$179 & -0.34 & -0.23 & -0.12 & -0.23 & -0.16 & -0.29 & -0.32 & -0.26 \\
 & & & & & & & &  \\
2016$-$615 & -0.69 & -0.51 & -0.49 & -0.56 &       &       &       &       \\
2052$-$474 & -0.23 & -0.10 & -0.17 & -0.16 & -0.31 & -0.12 &  0.08 & -0.12 \\
2058$-$297 & -0.25 & -0.03 &  0.54 &  0.07 & -0.20 &  0.04 &  0.10 & -0.02 \\
2106$-$413 &  0.23 &  0.25 &  0.09 &  0.20 &  0.10 &  0.31 &  0.11 &  0.18 \\
2109$-$811 &       &       &       &       & -0.03 &  0.18 &  0.27 &  0.14 \\
 & & & & & & & &  \\
2121+053   & -0.33 & -0.10 &  0.01 & -0.14 & -0.27 & -0.02 &  0.14 & -0.05 \\
2126$-$158 & -0.35 &  0.40 &  0.89 &  0.31 & -0.36 &  0.38 &  0.86 &  0.28 \\
2128$-$123 &  0.17 &  0.30 &  0.32 &  0.27 &  0.25 &  0.34 &  0.33 &  0.31 \\
2134+004   &       &       &       &       & -0.23 &  0.42 &  1.20 &  0.45 \\
2142$-$758 & -0.36 & -0.27 & -0.27 & -0.30 & -0.37 & -0.25 & -0.21 & -0.28 \\
 & & & & & & & &  \\
2146$-$783 & -0.51 &  0.23 &  0.68 &  0.13 & -0.53 &  0.21 &  0.61 &  0.09 \\
2149$-$307 & -0.22 &  0.29 &  0.33 &  0.14 & -0.34 &  0.20 &  0.33 &  0.06 \\
2155$-$304 & -0.03 &  0.10 &  0.07 &  0.05 & -0.04 &  0.08 & -0.03 &  0.01 \\
2243$-$123 &  0.05 &  0.26 &  0.17 &  0.17 &  0.05 &  0.24 &  0.14 &  0.15 \\
2245$-$328 &  0.07 & -0.16 & -0.41 & -0.16 &       &       &       &       \\
 & & & & & & & &  \\
2312$-$319 & -0.46 & -0.35 & -0.28 & -0.36 & -0.50 & -0.38 & -0.24 & -0.38 \\
2320$-$035 & -0.09 & -0.02 & -0.09 & -0.06 & -0.16 &  0.01 & -0.02 & -0.05 \\
2326$-$477 & -0.07 & -0.29 & -0.40 & -0.25 & -0.09 & -0.29 & -0.36 & -0.25 \\
2329$-$384 & -0.48 & -0.01 &  0.32 & -0.06 &       &       &       &       \\
2332$-$017 & -0.25 & -0.11 & -0.02 & -0.13 & -0.30 & -0.13 &  0.07 & -0.12 \\
 & & & & & & & &  \\
2333$-$528 & -0.29 & -0.16 & -0.19 & -0.21 & -0.33 & -0.16 &  0.05 &  0.18 \\
2345$-$167 & -0.16 &  0.29 &  0.01 &  0.06 & -0.07 &  0.20 &  0.18 &  0.11 \\
2355$-$534 & -0.25 &  0.18 &  0.04 &  0.00 & -0.14 &  0.04 &  0.06 & -0.01 \\
\end{supertabular}}
\end{center}
\onecolumn
\newpage
\begin{table}
\caption[The list of sources which show modulation indices exceeding $3\sigma$ for any of the frequencies, 2.4, 4.8 and 8.6 GHz in any of the two observing sessions: May ($\mu_{M}$) and August 1994 ($\mu_{A}$).]{The list of sources which show modulation indices exceeding $3\sigma$ for any of the frequencies, 2.4, 4.8 and 8.6 GHz, in any of the two observing sessions: May ($\mu_{M}$) and August 1994 ($\mu_{A}$). Entrees in bold are those specifically $>3\sigma $.}
\label{tab:11}
\begin{center}
\begin{tabular}{c@{\hspace{2mm}}r@{\hspace{2mm}}r@{\hspace{2mm}}r@{\hspace{2mm}}r@{\hspace{2mm}}r@{\hspace{2mm}}r@{\hspace{2mm}}} \hline
 Source & \multicolumn{2}{c}{8.6 GHz} & \multicolumn{2}{c}{4.8 GHz} & \multicolumn{2}{c}{2.4 GHz} \\ 
  name & $\mu_{M}$ & $\mu_{A}$ & $\mu_{M}$ & $\mu_{A}$ & $\mu_{M}$ & $\mu_{A}$ \\  \hline
0104$-$408  & 0.011  & 0.003   & 0.013   & 0.004   &{\bf 0.044}   & 0.007   \\
0346$-$279  & 0.009  & 0.007   & {\bf 0.030}   & 0.015   & 0.016   & 0.022   \\
0405$-$385  & {\bf 0.043}  & 0.003   & {\bf 0.056}   & 0.007   & {\bf 0.056}   & 0.011   \\
0422+004    & 0.018  &         & {\bf 0.024}   &         &         &         \\  
0437$-$454  & {\bf 0.025}  & 0.009   & 0.016   & 0.014   &         &        \\   
0440$-$003  &        &{\bf 0.029}   &         & 0.011   &         & 0.018    \\
0450$-$469  & 0.012  & 0.003   & 0.016   & 0.004   &{\bf 0.034}   & 0.006   \\
0502+049    & 0.005  & {\bf0.027}   & 0.018   & 0.011   &         &          \\
0528$-$250  & {\bf 0.020}  & 0.009   & 0.014   & 0.005   &         &          \\    
0607$-$157  & 0.012  & 0.005   & 0.013   & {\bf 0.023}   & 0.013   & {\bf 0.031}   \\
0646$-$306  & 0.014  & 0.012   & {\bf 0.024}   & 0.011   & {\bf 0.037}   & 0.024    \\
0728$-$320  &{\bf  0.026}  & 0.011   & {\bf 0.034}   & 0.007   &         &         \\
0808+019    & 0.014  & {\bf 0.021}   & 0.011   & {\bf 0.023}   & {\bf 0.025}   & 0.022    \\
1034$-$293  & {\bf 0.057}  & 0.012   & {\bf 0.092}   & {\bf 0.113}   &         &         \\                                              
1048$-$313  &{\bf 0.026}  & 0.019   & 0.011   & {\bf 0.025}   & 0.023   & 0.010    \\
1144$-$379  & {\bf 0.075}  & 0.017   & {\bf 0.146}   & {\bf 0.031}   &         &      \\
1256$-$220  & 0.012  & {\bf 0.027}   & 0.014   & {\bf 0.022}   & 0.009   & 0.007   \\  
1502+036    & {\bf 0.020}  & 0.002   & 0.014   & 0.008   &         &         \\  
1519$-$273  & {\bf 0.029}  & 0.018   & {\bf 0.037}   & {\bf 0.023}   & {\bf 0.040}   & 0.009   \\
1556$-$245  & 0.015  & {\bf 0.042}   & {\bf 0.028}   & {\bf 0.033}   &         &        \\        
1622$-$297  & {\bf 0.020}  & {\bf 0.031}   & 0.011   & 0.014   & 0.021   & 0.016    \\
2155$-$304  & {\bf 0.024}  & 0.008   & {\bf 0.021}   & 0.005   & {\bf 0.034}   & 0.020  \\
 \hline\end{tabular}
\end{center}
\end{table}
\newpage
\begin{table*}[t] 
\caption{The list of sources which show day-to-day polarized flux densities fluctuations larger than $3\sigma$ in any of the two observing sessions: May ($\Delta P_{M}$) and August 1994 ($\Delta P_{A}$).}
\begin{center}
\begin{tabular}{c@{\hspace{2mm}}r@{\hspace{2mm}}r@{\hspace{2mm}}r@{\hspace{2mm}}r@{\hspace{2mm}}r@{\hspace{2mm}}r@{\hspace{2mm}}} \hline
 Source & \multicolumn{2}{c}{8.6 GHz} & \multicolumn{2}{c}{4.8 GHz} & \multicolumn{2}{c}{2.4 GHz} \\ 
  name & $\Delta P_{M}$ & $\Delta P_{A}$ & $\Delta P_{M}$ & $\Delta P_{A}$ & $\Delta P_{M}$ & $\Delta P_{A}$ \\  \hline
0150$-$334  &  -0.006 & {\bf -0.009} & -0.002 & 0.005 & -0.003 & -0.004 \\
0727$-$115  & {\bf 0.009}  &     & -0.006   &     & {\bf 0.011}   &     \\
1334$-$127  & 0.000 &  & {\bf 0.010}  &  &  0.001 &  \\
1504$-$166  & 0.002 &  0.003 & -0.001 & 0.007 & {\bf 0.010} & {\bf 0.011} \\
1622$-$297  &       & 0.003  &        & {\bf 0.072} &       & {\bf -0.085}  \\
2106$-$413  & 0.004 & {\bf 0.020} & 0.001 & 0.000 &  -0.002 & -0.001   \\
2149$-$307  &       & -0.001 &  & 0.003 &   & {\bf  0.015} \\
2155$-$304  &       & {\bf 0.024}  &   & -0.005 &   & {\bf  0.010} \\     
2312$-$319  &   & {\bf -0.011}   &    & {\bf 0.009}   &    & {\bf 0.016}     \\
 \hline 
\end{tabular}
\label{tab:150}
\end{center}
\end{table*}

\newpage
\newpage
\newpage
\newpage
\newpage
\newpage
\newpage
\end{document}